\documentclass[aps,prb,twocolumn,floatfix,showpacs,superscriptaddress]{revtex4-1}
\usepackage{color,graphicx}
\usepackage{epsfig}
\usepackage{subfigure}
\usepackage{amsmath,amsfonts,amssymb,amsthm,verbatim}
\usepackage{tabularx}
\usepackage{bm}

\def\be{\begin{equation}}
\def\ee{\end{equation}}
\def\ba{\begin{eqnarray}}
\def\ea{\end{eqnarray}}
\def\Cal{\mathcal}

\begin{document}
\title{Interaction-driven topological and nematic phases on the Lieb lattice}
\author{Wei-Feng Tsai}
\email{wftsai@mail.nsysu.edu.tw}
\affiliation{Department of Physics, National Sun Yat-sen University, Kaohsiung 804, Taiwan, R.O.C.}
\author{Chen Fang}
\email{chen.fang2491@gmail.com}
\affiliation{Department of Physics, Massachusetts Institute of Technology, Cambridge, MA 02139, U.S.A.}
\author{Hong Yao}
\email{yaohong@tsinghua.edu.cn}
\affiliation{Institute for Advanced Study, Tsinghua University, Beijing 100084, China}
\author{Jiangping Hu}
\email{hu4@physics.purdue.edu}
\affiliation{Beijing National Laboratory for Condensed Matter Physics and
Institute of Physics, Chinese Academy of Sciences, Beijing 100080, China}
\affiliation{Department of Physics, Purdue University, West Lafayette, IN 47907, U.S.A.}
\date{\today}
\newcommand{\br}{\mathbf{r}}
\newcommand{\brprime}{{\mathbf{r}^\prime}}
\newcommand{\bk}{\mathbf{k}}
\newcommand{\bkprime}{{\mathbf{k}^\prime}}

\begin{abstract}
We show that topological states are often developed in two-dimensional semimetals with
quadratic band crossing points (BCPs) by electron-electron interactions. To illustrate this, we construct a concrete model with the BCP on an extended Lieb lattice and investigate the interaction-driven topological instabilities. We find that the BCP is marginally unstable against infinitesimal repulsions. Depending on the interaction strengths, topological quantum anomalous/spin Hall, charge nematic, and nematic-spin-nematic phases, develop separately. Possible physical realizations of quadratic BCPs are provided.
\end{abstract}

\maketitle

\section{Introduction}
The search of new topological states of matter has never stopped since the discovery of the quantum Hall state in 1980's.\cite{klitzing86} In particular, in recent years the study of ``topological insulators'' (TI) becomes one of the most active fields in condensed-matter physics, not only for its importance to fundamental physics but also for its potential application in spintronics or thermoelectrics.\cite{moore10,rev} This new insulating phase is distinguished from the conventional one by a non-trivial Z$_2$ topological invariant and robust gapless edge states in two dimensions (2D)\cite{kane05a,kane05b,bernevig06a,bernevig06b} or surface states in three dimensions (3D)\cite{fu07a,fu07b}, against moderate perturbations which preserve time reversal symmetry (TRS).

Generally, such topological insulating state can occur in a system with strong spin-orbit coupling that explicitly breaks spin rotational symmetry (SRS), resulting in the band inversion phenomenon. Typical materials which exhibit TI phase are found in, for instance, the HgTe/CdTe quantum wells (2D), Bi$_{x}$Sb$_{1-x}$, Bi$_2$Se$_3$, Bi$_2$Te$_3$ (3D), and so on.\cite{koenig07,hsieh08,hsieh09,xia09,chen09} However, an alternative route to TI is possible and it is associated with the concept of topological Mott insulator, first 
studied in Ref. [\onlinecite{scz08}] 
in {\it strongly correlated} systems. There are two key and generic ingredients in this approach. First, the Fermi surface of the non-interacting system should shrink to discrete points (2D) or lines (3D), and hence it is semi-metallic; second, there exists a suitable repulsive interaction, which can dynamically generate spin-orbit coupling through spontaneously broken SRS, as first discovered by Wu and Zhang.\cite{wu04} A few pioneering examples along this line of thought have been discussed in various lattice geometries, {\it e.g.}, honeycomb,\cite{scz08,weeks10a,xu10,yingran,fiete2,fiete3,fiete4} checkerboard,\cite{sun09} kagome,\cite{liu10,fiete1}, diamond lattices\cite{zhang09}, and in the low-energy continuum theory\cite{oskar14}.

\begin{figure}[tbh]
\begin{center}
\includegraphics[scale=0.5]{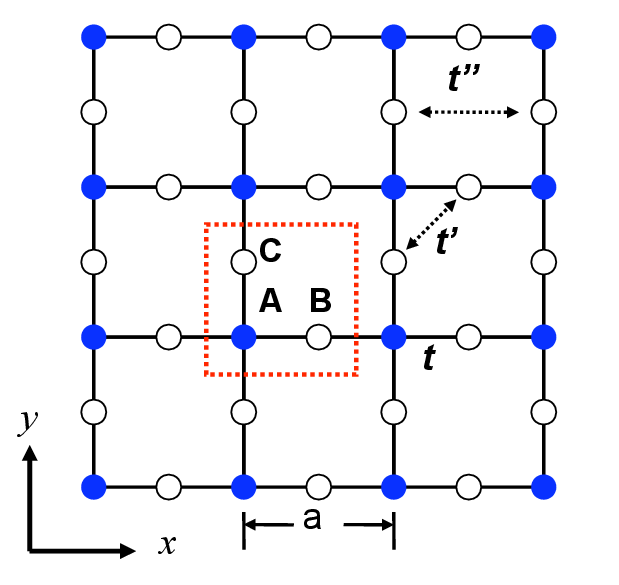}
\caption{(Color online) Schematic plot of the 2D (extended) Lieb lattice. The
dotted curve indicates the unit cell, which contains three
sublattice sites (A, B, C).}
\label{fig:lattice}
\end{center}
\end{figure}

Here we show that topological states can be generally realized in a system with quadratic band crossing points (BCPs), which are symmetry protected at non-interacting level. Near such kind of BCP, instability towards phases with broken symmetries
is inevitable even if there is only weak interaction between electrons. To demonstrate this, we construct a concrete model with such a BCP on an extended 2D Lieb lattice. There are several reasons for us to choose the Lieb lattice. First, the Lieb lattice has three sites per unit cell, as shown in Fig.~\ref{fig:lattice}. With only nearest-neighbor (NN) hoppings, there is a {\it dispersionless} (flat) band in the middle of the band structure. The three-band touching point is the result of a combined effect of (a) crystalline symmetry group of the Lieb lattice ($D_{4h}$-group), (b) spin rotation symmetry (or the absence of spin-orbital coupling) and (c) equal intra-sublattice hopping amplitudes and vanishing hopping amplitudes between B- and C-sublattices.
Depending on the values of the onsite potentials or the presence of further range hoppings, the band crossing feature between this flat band and the two other bands may include:
(i) The flat band touches upper and lower linearly dispersing bands at one point when (a,b,c) are all satisfied. (ii) The flat band can be isolated. For instance, one can add intrinsic spin-orbit couplings, i.e., breaking (b), as discussed by Weeks and Franz.\cite{weeks10b} (iii) When the onsite potential on $A$ sublattice is not equal to that of $B$/$C$ sublattices [i.e., breaking (c)], the flat band only touches one of the other two bands, which becomes quadratically dispersing, instead of linearly dispersing.\cite{green10} Thus, these choices could variegate our results. Second, a nearly flat band has effectively large correlation effects due to the small bandwidth, leading to fractional Chern insulating phases if the band has nonzero Chern number (see Sec. IV). Finally, the 2D Lieb lattice has been the most important building block in many 3D perovskite materials featured with complex phase diagram and strong electron-electron correlations. Thus study of the model can be viewed as a preliminary investigation of the TI phase especially in layered perovskites composed of weakly coupled 2D planes with Lieb lattice structure ({\it e.g.}, the well-known high-$T_c$ cuprates).

In this paper, we start with the construction of the explicit model and reveal the topological nature of the BCP at non-interacting level. We then examine the consequence of such topological BCP, {\it i.e.}, with a symmetry protected quadratic dispersion, under the presence of short-range repulsive interactions. We investigate various symmetry breaking instabilities at BCP within self-consistent mean-field approximation. Note that we mainly focus on type 
(iii) band structure, namely, only two bands touch together, and compare it with the case of type (i) when necessary. In principle, for the BCP there are two ways to open a gap and gain energy: One is to open a full gap at BCP, and the other one is to split the BCP into two Dirac points (each with Berry flux $\pi$), but at the price of broken $C_4$ symmetry. To justify this speculation, we show phase diagrams for spinless/spinful fermions at 1/3 or 2/3 filling, according to the position of the BCP in the band structure. In fact, at both fillings the phase diagrams are qualitatively similar with subtle differences due to particle-hole asymmetry introduced by the interactions. In the spinless case, the leading order under ``weak'' next nearest-neighbor (NNN) repulsion is the quantum anomalous Hall (QAH) insulating state (TRS broken). For ``strong'' NNN repulsion, the ground state evolves into insulating nematic state ($C_4$ symmetry broken down to $C_2$). In addition, for intermediate strength, there exists a narrow coexistence region between these two orders. In the spinful case, the phase diagrams are more complicated. Besides the phases we find in the spinless case, there are also a quantum spin Hall (QSH) insulating state and a nematic-spin-nematic semi-metallic phase with Dirac nodes.\cite{oganesyan01,kivelson03,wu07}
Thus, we clearly demonstrate that, in principle, correlated systems with Lieb lattice structure can be a {host to various nontrivial phases including TIs}.

\begin{figure}[tbh]
\def\subfigcapskip{0pt}
\def\subfigtopskip{5pt}
\def\subfigbottomskip{5pt}
\subfigure[]{\includegraphics[width=0.35\textwidth]{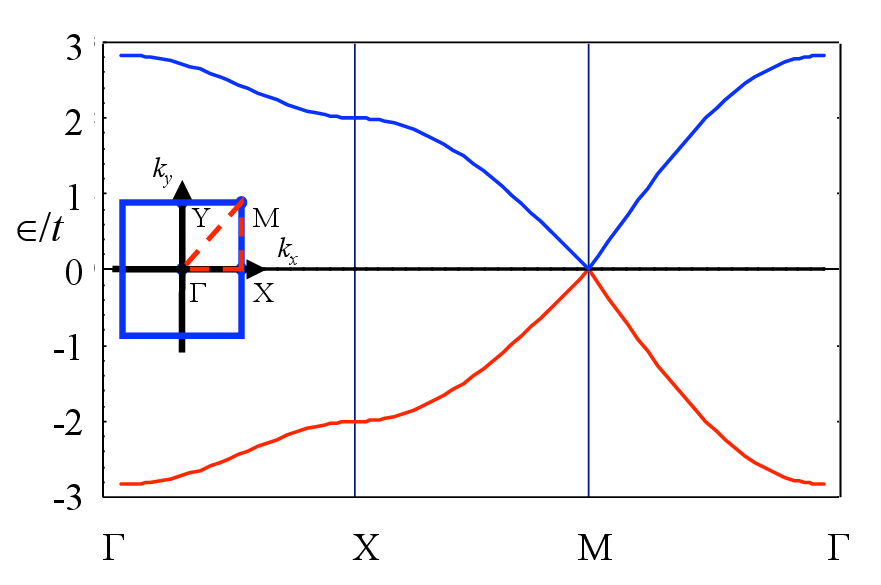}}
\subfigure[]{\includegraphics[width=0.38\textwidth]{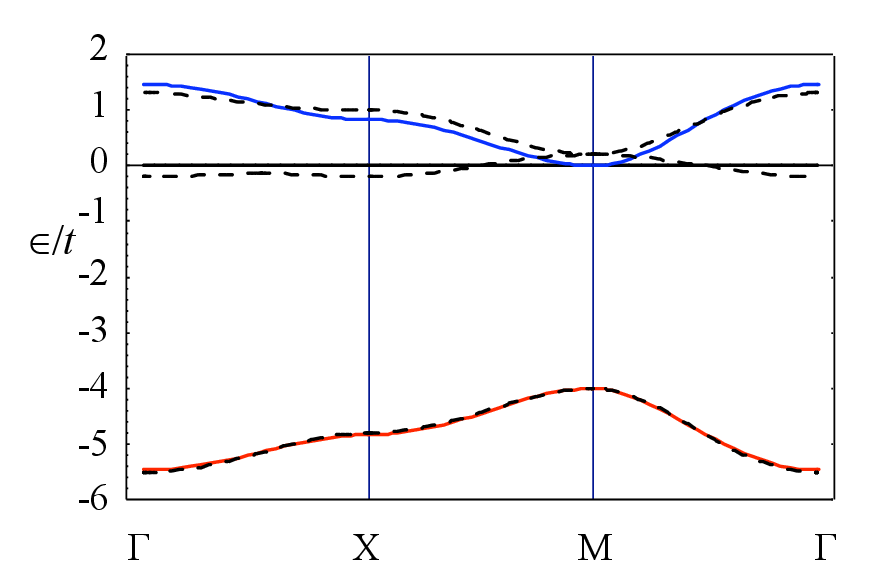}}
\caption{(Color online) Band structure of the model along the high-symmetry lines in the FBZ for (a) $\varepsilon_A=0$ and for (b) $\varepsilon_A=-4t$. In case (b), when $t^{\prime\prime}=0.1t$ (with the constraint mentioned in the text) is present, the BCP at $\mathbf{M}$ becomes standard QBCP and the spectrum is changed slightly as shown by dashed lines. Inset of (a): FBZ of the Lieb lattice. There are four time reversal invariant momenta: $\mathbf{\Gamma},\mathbf{X},\mathbf{M}$, and $\mathbf{Y}$.}
\label{fig:spectrum}
\end{figure}

This paper is organized as follows. In Section II, we define the model and demonstrate the topological nature of its BCP from both momentum and real space points of view. Next, the consequences of introducing short-range repulsions are discussed for spinless fermions in Section III A and for spinful fermions in Section III B, respectively. Finally, we discuss some issues and make conclusions in Sec IV.

\section{The lattice model}
We begin with the tight-binding model for non-interacting fermions,
\be
\Cal{H}_0=-\sum_{ij}t_{ij}c^\dagger_{i}c_{j}
+\sum_{i}\varepsilon_i c^\dagger_{i}c_{i},
\label{eq:H0}
\ee
where $c^\dagger_{i}$ creates a fermion on site $i$ of the 2D Lieb lattice, the unit cell of which is given by $A,B,C$ sites shown in Fig.~\ref{fig:lattice}. For simplicity, we take the hopping amplitudes, $t_{ij}=t$, between NN sites $\langle ij \rangle$, and $t_{ij}=0$ otherwise for the moment. The effect of adding longer-range hopping amplitudes (but small in magnitude) will be discussed later when appropriate. 
Note that the $C_4$ point group symmetry dictates that the onsite potentials on the $B$ and $C$ are equal, $\varepsilon_B=\varepsilon_C$. 
Although $\varepsilon_{(A,B,C)}$ are generically non-zero, only their
relative values are essential to determine the symmetry of the
lattice, and hence, the band structure. Therefore, hereafter we set
the units of energy $t\equiv 1$, the lattice constant $a\equiv 1$, and, without loss of generality, $\varepsilon_B=\varepsilon_C=0$.

\subsection{Band structure}
The band structure of Eq.~(\ref{eq:H0}) can be obtained by
transforming $\Cal{H}_0$ into momentum space,
\be
\Cal{H}_0=\sum_{\bk}\psi^\dagger_{\bk}H_0(\bk)\psi_{\bk}, \label{eq:h0}
\ee
where the fermion spinor, $\psi^\dagger_{\bk}=(c^\dagger_{A\bk},c^\dagger_{B\bk},
c^\dagger_{C\bk})$, with sublattice (basis) index $A,B,C$ and $\bk=(k_x,k_y)$. Defining the displacement vectors, $\mathbf{a}_1=(1/2,0)$ and $\mathbf{a}_2=(0,1/2)$, $H_0(\bk)$ is of the form
\be
H_{0}(\bk)=\left(
\begin{array}
[c]{ccc}%
\varepsilon_A & -2t\cos\left(  \mathbf{k}\cdot\mathbf{a}_{1}\right)  & -2t\cos\left(
\mathbf{k}\cdot\mathbf{a}_{2}\right) \\
                      & 0 & 0 \\
                      &    & 0
\end{array}
\right),
\ee
where the lower triangular matrix is understood to be filled for keeping
the whole matrix hermitian. In this notation, the first Brillouin zone
(FBZ) is a square with four time reversal invariant momenta (TRIM):
$\mathbf{\Gamma}=(0,0),\mathbf{X}=(\pi,0),\mathbf{M}=(\pi,\pi)$, and
$\mathbf{Y}=(0,\pi)$ [see the inset of Fig.~\ref{fig:spectrum}(a)].
The energy spectrum consists of two dispersive bands,
$\epsilon_{\pm}(\bk)=\frac{1}{2}(\varepsilon_A\pm\sqrt{\varepsilon_A^2+4b_{\bk}})$ with
$b_{\bk}=\sum_{i=1}^{2}[2t\cos(\bk\cdot \mathbf{a}_i)]^2$, and one
{\it dispersionless} flat band, $\epsilon_0(\bk)=0$.

An important feature of this model is that the presence or absence of
$\varepsilon_A$ can change electronic properties dramatically. When
$\varepsilon_A=0$, the flat band touches two linearly dispersing bands
at $\mathbf{M}$ point in the FBZ [type (i)], where the linear bands meet as if there were a ``Dirac point''. However, the touching point in fact has completely different structure. It becomes clear once we expand $H_0(\bk)$ around
$\mathbf{M}$ point with $\bk=\mathbf{M}+\mathbf{p}, |\mathbf{p}|\ll
1$. To the first order in $\mathbf{p}$, $H_0(\bk)$ can be written as
\be
H_0(\mathbf{B})\sim v_F \mathbf{L}\cdot \mathbf{B},
\label{eq:spin1h0}
\ee
where the Fermi velocity $v_F=t$, $\mathbf{B}=(p_x,p_y,0)$, and the
(pseudo) spin-1 matrices are defined as
\be
L_x=\left(
\begin{array}[c]{ccc}
0 & 1 & 0 \\
1 & 0 & 0 \\
0 & 0 & 0
\end{array}
\right),
L_y=\left(
\begin{array}[c]{ccc}
0 & 0 & 1 \\
0 & 0 & 0 \\
1 & 0 & 0
\end{array}
\right),
L_z=\left(
\begin{array}[c]{ccc}
0 & 0 & 0 \\
0 & 0 & -i \\
0 & i & 0
\end{array}
\right),
\ee
obeying Lie algebra of SU(2), i.e.,
$[L_i,L_j]=i\epsilon_{ijk}L_k$,\cite{goldman11,shen11} instead of a Clifford
algebra as in the case of graphene. This is the fundamental reason why
there is no Dirac point and hence no fermion doubling problem\cite{nielssen83} on the Lieb lattice. Viewing the low-energy effective Hamiltonian $H_0(\mathbf{B})$ as a ``spin'' $\mathbf{L}$ in an external ``magnetic field'' $\mathbf{B}$, its eigenvalues can be easily read out as $\epsilon(\mathbf{B})=v_F|\mathbf{p}|l_{\mathbf{p}}$, where
$l_{\mathbf{p}}=0,\pm 1$ are the quantized angular momenta along the axis parallel to $\mathbf{B}$ in three dimensions.

\begin{figure}[tbh]
\includegraphics[scale=0.32]{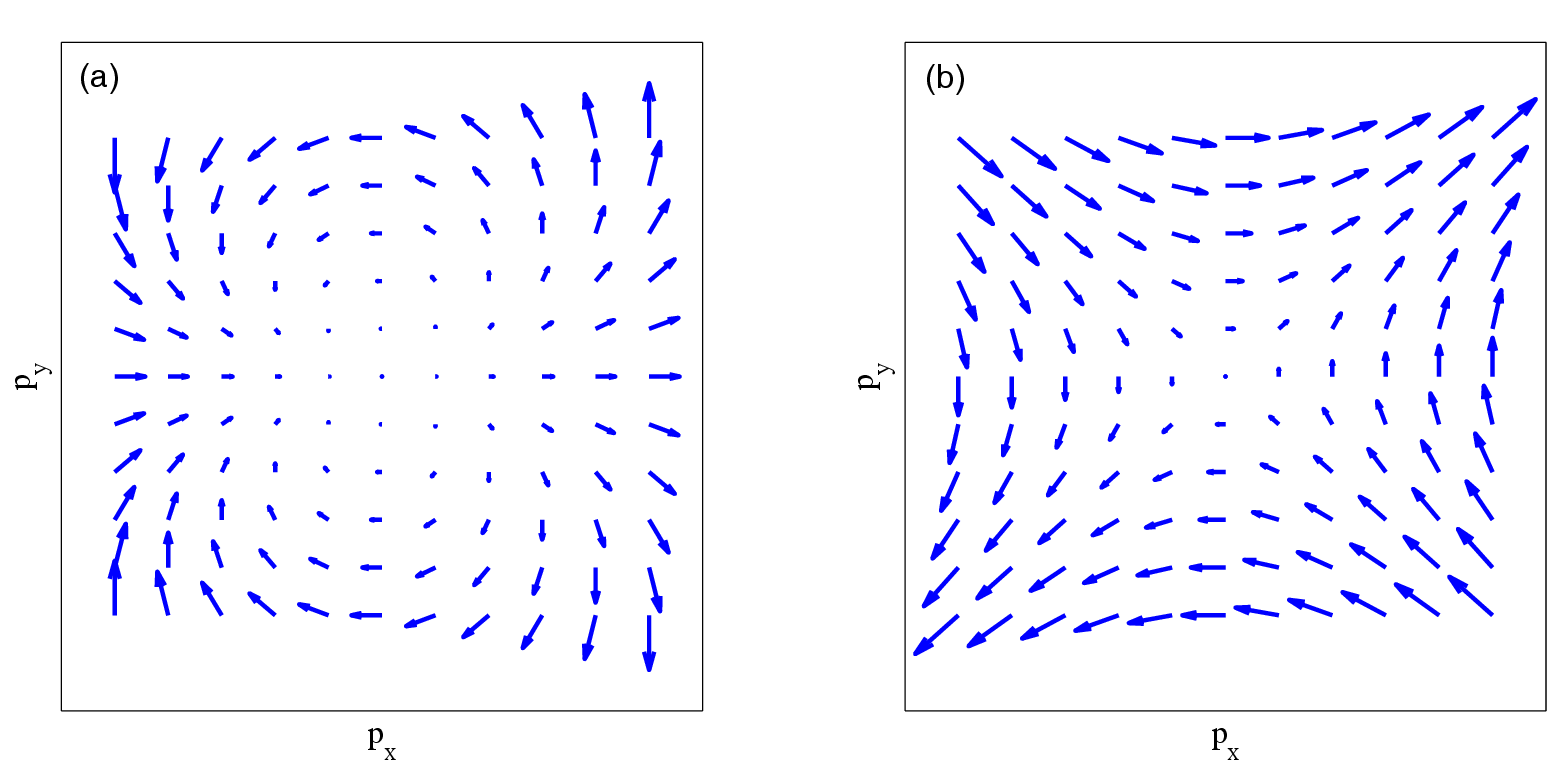}
\caption{(Color online) Vortex structure of the BCP in $\mathbf{p}$ space: (a) the planar vector field $\mathbf{B}^\prime$; (b) the planar vector field $\mathbf{B}$.}
\label{fig:vortex}
\end{figure}

When $\varepsilon_A\neq 0$, however, the spin-1 structure mentioned
above is no longer valid. The flat band now touches only one
dispersive (massive) energy band either above or below at $\mathbf{M}$ point, depending on the sign of
$\varepsilon_A$ [see Fig.~\ref{fig:spectrum} (b)]. To make this
structure transparent, we again expand $H_0(\bk)$ around
$\mathbf{M}$ with small $\mathbf{p}$. Assuming $|\mathbf{p}|\ll|\varepsilon_A/t|$
and $\varepsilon_A<0$ at 2/3 filling, we then integrate out the contribution from
basis $A$ (due to almost fully filled $A$ sublattice) and obtain a
low-energy effective two-band Hamiltonian,
\be
H^{eff}_0(\mathbf{p})\sim \frac{1}{m_0}\left(
\begin{array}[c]{cc}
p_x^2 & p_x p_y  \\
p_x p_y & p_y^2
\end{array}
\right)
=d_I I+d_x\sigma_x+d_z\sigma_z,
\label{eq:heff}
\ee
where $m_0=-\varepsilon_A/v_F^2$. In the last equality, we
express $H_0^{eff}$ in terms of the identity and Pauli matrices with
$d_I=\frac{1}{2}(p_x^2+p_y^2)m_0^{-1}$, $d_x=p_x p_y/m_0$, and
$d_z=\frac{1}{2}(p_x^2-p_y^2)m_0^{-1}$. Interestingly, if we further allow small, lattice symmetry unbroken third-neighbor hoppings $t^{\prime\prime}>0$ (but {\it forbid} to hop whenever here is a site in the middle of the path), the flat band becomes
slightly dispersive and the effective Hamiltonian changes to $d_I=\frac{1}{2}(p_x^2+p_y^2)(m_0^{-1}-t^{\prime\prime})$, $d_x=p_x p_y/m_0$, and
$d_z=\frac{1}{2}(p_x^2-p_y^2)(m_0^{-1}+t^{\prime\prime})$ without
removing the band crossing point (BCP) [see Fig.~\ref{fig:spectrum} (b)].
Such point at $\mathbf{p}=0$ is the
so-called quadratic band crossing point (QBCP), which has been studied
recently by several research groups.\cite{chong08,sun08,
  sun09,uebelacker11}
One of the key features for QBCP in 2D is that its density of states
(DOS) is non-zero at the crossing point, in sharp contrast to the case of Dirac points. This will
lead to essential difference when responding to the weak interactions
present in the system. In the following, we will mainly focus on the $\varepsilon_{A}\neq 0$ case and show that the BCP in our model is not only topologically non-trivial, but also make the system be a potential host to a topological phase under weak repulsive interactions.

\subsection{Topological nature of the band touching}
The band touching phenomenon on the Lieb lattice is quite generic and
stable for non-interacting fermions. Such stability deserves a full
analysis here. We shall provide two different approaches to show it:
One is based on momentum-space topology, and the other one is based on
real-space topology.

From the first point of view, the BCP actually forms a topological defect in the momentum space, similar to a vortex in a 2D superconductor, here with a winding number $\pm 2$. To see this, let us rewrite Eq.~(\ref{eq:heff}) as $H^{eff}_0(\mathbf{p})=d_I I+\mathbf{B}^\prime(\mathbf{p})\cdot\mathbf{\sigma}$, where the ``magnetic field'' $\mathbf{B}^\prime(\mathbf{p})=(d_x,0,d_z)$. This effective Hamiltonian now represents a spin-1/2 particle sitting in a magnetic field $\mathbf{B}^\prime$, which has a vortex structure at $\mathbf{p}=0$, as shown in Fig.~\ref{fig:vortex}(a). For comparison, recall that for $\varepsilon_A=0$ we instead have a spin-1 particle in an external field $\mathbf{B}$ [Eq.~(\ref{eq:spin1h0})], whose structure is shown in Fig.~\ref{fig:vortex}(b). The winding number $W$ then can be easily extracted from the figures that in the former case, $W=2$; in the latter case, $W=1$. However, somewhat counter intuitively, both cases are associated with the same Berry phase of the BCP,\cite{haldane04} which can be calculated precisely by
\be
B^{n}=i\oint_{\Gamma}d\mathbf{p}\cdot\langle u_{n\mathbf{p}}|\nabla_\mathbf{p}|u_{n\mathbf{p}}\rangle,
\label{eq:berry}
\ee
where $\Gamma$ is a contour in the $\mathbf{p}$ space enclosing the touching point, $n$ denotes any one of the involved bands, and $|u_{n\mathbf{p}}\rangle$ represents the Bloch wave function for $n$th band. A simple argument solves this puzzle. The line integral along any loop enclosing 0 in $\mathbf{p}$ space given above is known to be 1/2 (1) times the solid angle subtended by $\mathbf{B}^\prime(\mathbf{p})$ [$\mathbf{B}(\mathbf{p})$] from the origin for a spin 1/2 (1) particle. Thus, $B^n=2\pi W/2= 2\pi$ in the former case, which is just equivalent to $B^n=2\pi W= 2\pi$ in the latter one. In fact, when $B^n =0$, any infinitesimal mixing (perturbation) between bands would lift the degeneracy. With non-vanishing $B^n=\pm 2\pi$, we confirm that the BCP on the Lieb lattice at non-interacting level is topologically stable (i.e., not opening a gap) as long as the spinless system preserves both TRS and $C_4$ point group symmetry. Note that $C_4$ symmetry in our model is quite essential, as a similar QBCP happens in the A-B stacking bilayer graphene (with $C_3$ symmetry) while it can easily decay into Dirac BCPs and thus is topologically unstable.\cite{sun09,vafek10}

\begin{figure}[tbh]
\begin{center}
\includegraphics[scale=0.35]{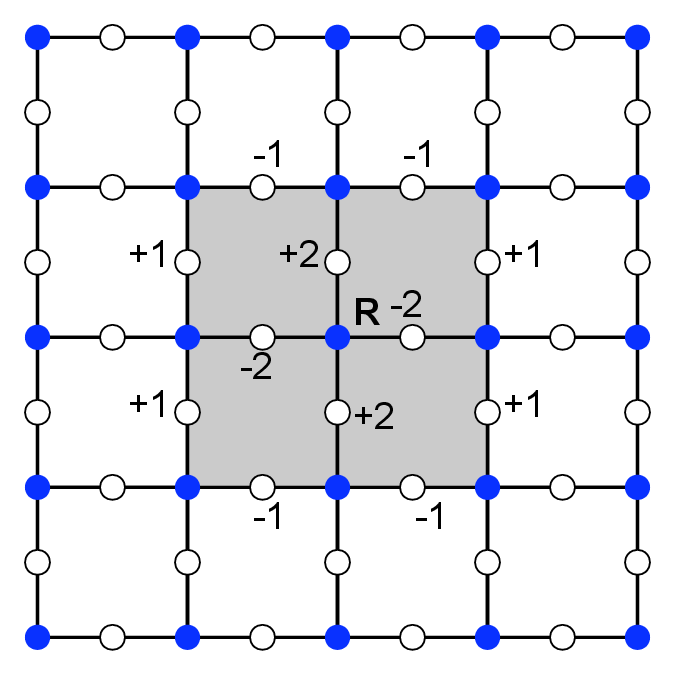}
\caption{(Color online) Schematic plot of the localized eigenstate at $\mathbf{R}$ on the Lieb lattice ($\varepsilon_A\neq 0$). Only those sites in the shaded area have non-zero weights, whose magnitudes are denoted by numbers (up to a normalization factor). The weights on $A$ sublattice (blue dots) are all zero.}
\label{fig:wannier}
\end{center}
\end{figure}

An alternative point of view for the protection of such BCP on the
Lieb lattice can be associated with certain topological structure
present in the real space, or more specifically, with the existence of the
eigenstates which are extended along {\it non-contractible} loops winding
around the whole lattice with periodic boundary conditions ({\it i.e.}, a
torus).\cite{bergman08} To demonstrate this feature, we first take the
merit of the flat band, which allows us to construct its corresponding
localized, one-particle eigenstates of $\Cal{H}_0$ (Wannier
states). Taking $\mathbf{R}$ to be the coordinate of the central
site of the shaded plaquette shown in Fig.~\ref{fig:wannier}, we find that the creation operator for the localized eigenstate at $\mathbf{R}$ can be written as
\be
\Cal{A}^\dagger_{\mathbf{R}}=\frac{1}{2\sqrt{6}}
[\sum_{j=1}^{4}(-1)^j(2c^\dagger_{\mathbf{R}+\mathbf{b}_j}
-c^\dagger_{\mathbf{R}+\mathbf{b}_{2j+3}}
-c^\dagger_{\mathbf{R}+\mathbf{b}_{2j+4}})], \label{eq:wannier}
\ee
where $\mathbf{b}_1=\mathbf{a}_1,\mathbf{b}_2=\mathbf{a}_2,
\mathbf{b}_3=-\mathbf{b}_1,\mathbf{b}_4=-\mathbf{b}_2$,
$\mathbf{b}_5=(1,-1/2),\mathbf{b}_6=(1,1/2),
\mathbf{b}_7=(1/2,1),\mathbf{b}_8=(-1/2,1)$,
$\mathbf{b}_9=-\mathbf{b}_5,\mathbf{b}_{10}=-\mathbf{b}_6,
\mathbf{b}_{11}=-\mathbf{b}_7,\mathbf{b}_{12}=-\mathbf{b}_8$.
The key reason for these states being localized is rooted on the fact that all A-sites have vanishing amplitudes and remain zero after the action of $\Cal{H}_0$ on them due to destructive interference.

\begin{figure}[bth]
\begin{center}
\includegraphics[scale=0.35]{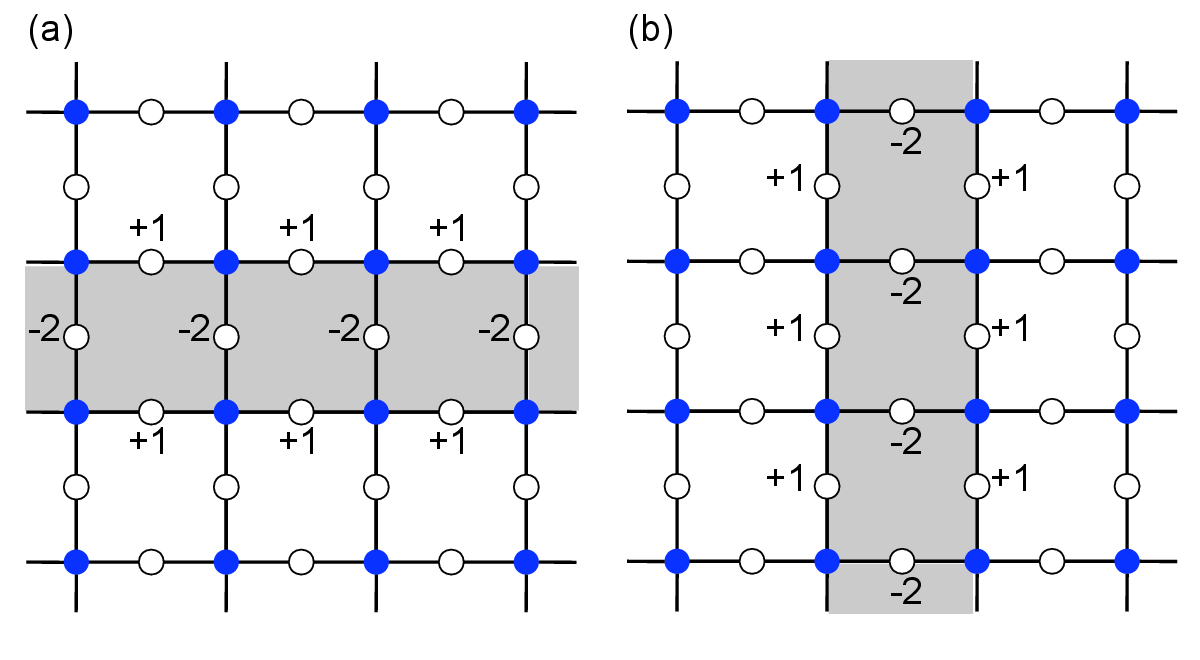}
\caption{(Color online) The two non-contractible loop states winding the lattice along periodic $x$ direction (a) and $y$ direction (b). Only those sites in the shaded area have non-zero weights (denoted by numbers in the plots). The weights on $A$ sublattice (blue dots) are all zero.}
\label{fig:ncloop}
\end{center}
\end{figure}

In localized-state language, the existence of the BCP in our model with $\varepsilon_A\neq 0$ is equivalent to state that the dimension of the space expanded by independent localized eigenstates with zero energy has a dimension which is one larger than the number of unit cells, $N$. The extra state cannot come from the flat band, but from one of the dispersive bands. The plaquette states we constructed in Eq.~(\ref{eq:wannier}) seem to form $N$ linearly independent states with zero energy. For our model with periodic boundary conditions, however, the following relation,
\be
\Cal{A}^\dagger_{\mathbf{q}=(\pi,\pi)}=\sum_{\mathbf{R}}
e^{i\mathbf{q}\cdot\mathbf{R}}\Cal{A}^\dagger_{\mathbf{R}}=0,
\ee
reduces the naive counting by one and hence only $N-1$ states are independent. The missing two states, in fact, are accounted for by two {\it non-contractible} loops around the whole lattice (torus), as illustrated in Figs.~\ref{fig:ncloop}(a) and (b). When $\Cal{H}_0$ acts on these states, the destructive interference again guarantees the zero eigenvalue. Now, in total we have $N+1$ independent states. Therefore, provided not destroying the flat band, such band touching phenomenon is protected by the topological character of the lattice.

\section{Interaction driven instabilities}
The existence of such symmetry/topology protected BCP on the Lieb lattice at the non-interacting level 
motivates us to further ask if it is 
stable in the presence of repulsive interactions. To see this, we will first examine whether generic short-range repulsions are relevant to this BCP from perturbative renormalization group (RG)
analysis, and next, if the interactions are relevant, we will investigate possible consequences of such instability, i.e. symmetry breaking phases, at mean-field (MF) level.

To perform RG analysis, we consider a continuum, spin-1/2 Hamiltonian, which can be obtained by projecting the original three-band interacting model onto an {\it effective two-band theory} near the BCP in the continuum and $|\varepsilon_A|\gg t$ limit,
\be
\Cal{H}^{eff}=\Cal{H}^{eff}_0+\Cal{H}^{eff}_{int},
\ee
where the free part $\Cal{H}_0$ reads as
\be
\Cal{H}^{eff}_0=\sum_{\sigma}\int d\mathbf{r}\Psi^\dagger_{\sigma}(\mathbf{r})H^{eff}_0
\Psi_{\sigma}(\mathbf{r}).
\ee
The subscript $\sigma$ denotes spin polarization and the fermion field $\Psi^\dagger_\sigma=(\psi^\dagger_{1\sigma},\psi^\dagger_{2\sigma})$ with 1,2 representing orbital (i.e., two touching bands) degrees of freedom. Note that in the momentum space the expression for $H^{eff}_0$ is given by Eq.~(\ref{eq:heff}), and is independent of $\sigma$. For simplicity, we take small $t^{\prime\prime}=-t^2/\varepsilon_A$ such that $d_I=0$, making the effective theory particle-hole symmetric. In fact, non-vanishing $d_I$ would not change our main conclusion, provided $m_0^{-1}> t^{\prime\prime}$. The projected interacting part includes only 1) intra-orbital and 2) inter-orbital contact interactions,
\ba
\Cal{H}^{eff}_{int}&=& \Cal{H}^{eff}_{1}+\Cal{H}^{eff}_{2},  \\
\Cal{H}^{eff}_1 &=& \sum_{a=1}^{2}u\int d\mathbf{r}\psi^\dagger_{a\uparrow}(\mathbf{r})
\psi^\dagger_{a\downarrow}(\mathbf{r})
\psi_{a\downarrow}(\mathbf{r})
\psi_{a\uparrow}(\mathbf{r}), \nonumber \\
\Cal{H}^{eff}_2 &=& \sum_{\sigma,\sigma^\prime}g_{\sigma\sigma^\prime}\int d\mathbf{r}
\psi^\dagger_{1\sigma}(\mathbf{r})\psi^\dagger_{2\sigma^\prime}(\mathbf{r})
\psi_{2\sigma^\prime}(\mathbf{r})
\psi_{1\sigma}(\mathbf{r}), \nonumber
\ea
where $u$ and $g_{\sigma\sigma^\prime}$ are intra-orbital and inter-orbital coupling parameters, respectively.

For the chemical potential $\mu=0$, the non-interacting system $\Cal{H}^{eff}_0$ leads to one Fermi point at $\mathbf{p}=0$ with non-vanishing DOS, instead of a Fermi surface. Setting the dimension $[p]=1$ and understanding the dynamical critical exponent $z=2$ due to quadratic dispersion, it is straightforward to see that the dimension $[\psi_{a\sigma}(\mathbf{r})]=1$; in the interacting part the coupling constants $[u]=[g_{\sigma\sigma^\prime}]=0$, implying that they are superficially {\it marginal} interactions. However, as shown in Appendix A, we find that they are generically marginally relevant and bring the system to the strong coupling regime. More explicitly, up to one-loop order, the coupled RG equations for the coupling parameters are
\ba
\frac{du}{dl} &=& (2\gamma-\alpha)u^2+2\gamma g_{\uparrow\uparrow}u-2\gamma g_{\uparrow\uparrow}g_{\uparrow\downarrow}, \nonumber \\
\frac{dg_{\uparrow\uparrow}}{dl} &=& \alpha g^2_{\uparrow\uparrow}-2\gamma g_{\uparrow\downarrow}u+\gamma g^2_{\uparrow\downarrow}, \nonumber \\
\frac{dg_{\uparrow\downarrow}}{dl} &=& (\alpha-2\gamma)g^2_{\uparrow\downarrow}-2\gamma g_{\uparrow\uparrow}u+2\gamma g_{\uparrow\uparrow}g_{\uparrow\downarrow},
\label{eq:RGeq}
\ea
where we have used the fact that $g_{\uparrow\uparrow}=g_{\downarrow\downarrow}$ and $g_{\uparrow\downarrow}=g_{\downarrow\uparrow}$. $l$ denotes the momentum rescaling $p\rightarrow pe^{-l}$ and the coefficients $\alpha=\frac{1}{2\pi^2|\bar{t}|}K(\sqrt{1-\bar{t}^{-2}})$ and $\gamma=\frac{\bar{t}^2E(1-\bar{t}^{-2})-K(1-\bar{t}^{-2})+|\bar{t}|
(E(1-\bar{t}^2)-K(1-\bar{t}^2))}{8\pi^2(-1+\bar{t}^2)|\bar{t}|}$, with $K(x)$ [$E(x)$], the elliptic function of first (second) kind and $\bar{t}=(m_0^{-1}+t^{\prime\prime})/m_0^{-1}$. In fact, no new fixed point (FP) is produced in this set of RG equations, except for the non-interacting one at which $u$ and $g_{\sigma\sigma^\prime}$ vanish. Moreover, we find that given generic bare coupling parameters ($u,g_{\sigma\sigma^\prime}>0$), at least one of them diverges first when reaching low enough energy scale. This indicates that the non-interacting FP is an {\it unstable} FP, which can drive the system to strong coupling regime in the presence of short-range repulsions. In addition, it is worth mentioning that by setting $g_{\uparrow\downarrow}=u=0$, we reduce the RG equation back to the spinless case,
\be
\frac{dg_{\uparrow\uparrow}}{dl}=\alpha g^2_{\uparrow\uparrow}
\ee
with $\alpha>0$, which is consistent with the work done 
in Ref. \onlinecite{sun09} and, importantly, it implies that short-range repulsions are again {\it marginally relevant}.

\begin{figure}[tbh]
\begin{center}
\includegraphics[scale=0.35]{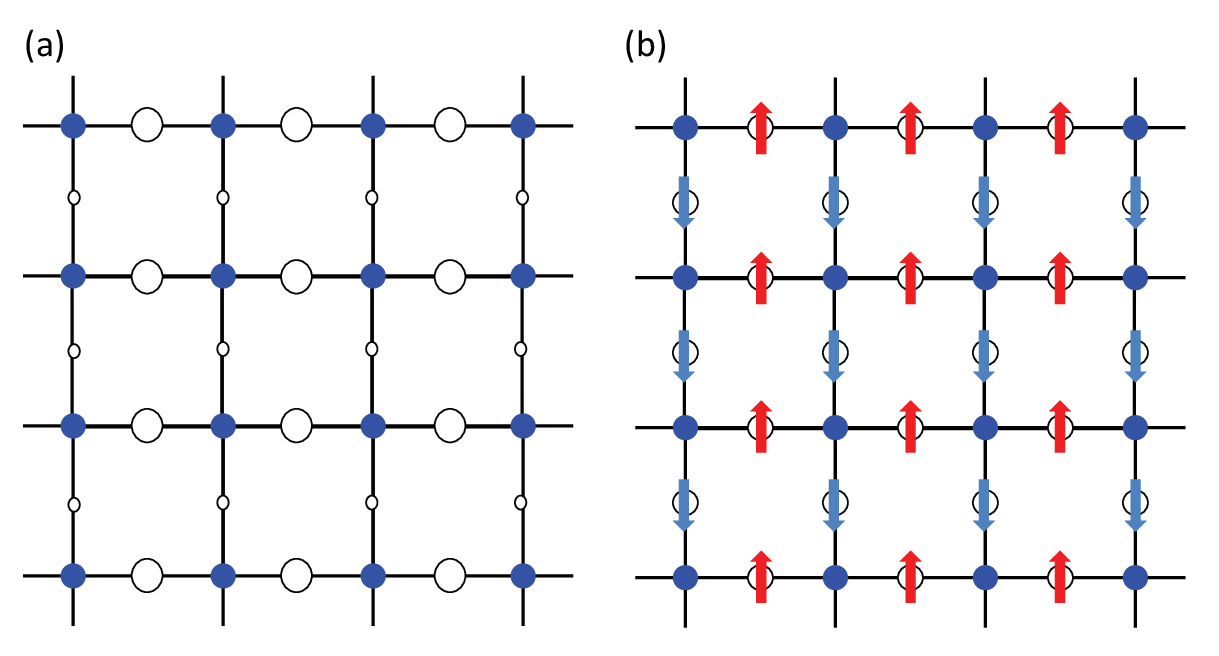}
\caption{(Color online) The schematic plots for (a) site charge nematic state and (b) site nematic-spin-nematic state. Since the electronic information of A sublattice sites (blue dots) is irrelevant here, only those of B and C sublattice sites are compared. Each open circle represents the local charge density, whose magnitude depends on the size of it. The arrow at each site represents a net spin polarization.}
\label{fig:nematic}
\end{center}
\end{figure}

\subsection{Spinless fermions}
From above we know that a QBCP is generally unstable against weak repulsive interactions. We now discuss its consequence on the Lieb lattice and explore possible symmetry breaking phases driven by interactions at mean-field level. We warm up with the spinless case to gain some physical insights before including spin degrees of freedom.

The lattice model we study is given by Eq.~(\ref{eq:H0}), with short-range interacting terms,
\be
\Cal{H}_{int}= V_1\sum_{\langle ij \rangle}n_i n_j+V_2\sum_{\langle\langle ij \rangle\rangle} n_i n_j,
\label{eq:int}
\ee
where $V_1$ and $V_2$ are repulsive coupling constants for NN and NNN interactions, respectively. $n_i=c^\dagger_i c_i$ is the number operator on the site $i$. The chemical potential is suitably chosen to keep the system at 2/3 (1/3) filling for $\varepsilon_A<0$ ($\varepsilon_A>0$). We proceed by treating $\Cal{H}_{int}$ in the MF approximation, including both the on-site and bond MF decoupling particle-hole channels,
\ba
n_i n_j &\rightarrow & n_i\langle n_j\rangle + n_j\langle n_i\rangle - \langle n_i\rangle\langle n_j\rangle, \\
n_i n_j &\rightarrow & -\phi_{ij}c^\dagger_j c_i - \phi^*_{ij} c^\dagger_i c_j +\phi_{ij}\phi^*_{ij},
\ea
where $\phi_{ij}=\phi^*_{ji}=\langle c^\dagger_i c_j\rangle$ represent certain current/bond order with $i,j$ belonging to NN and NNN bonds. Note that in this work only {\it translation-invariant} MF ansatz is considered. The repulsive interactions can produce the following possible phases:

(i) {\it Nematic state}. This is a phase associated with broken $C_4$ symmetry down to $C_2$. In particular, it does not break the translational symmetry by any lattice vectors, and thus is in contrast to the conventional charge density wave (CDW) order, which enlarges the unit cell due to translational symmetry breaking [See Fig.~\ref{fig:nematic}]. This phase behaves like an anisotropic metal (one QBCP splits into two Dirac points) or an insulator (two Dirac points meet at zone boundary and end up with a gap), depending on the strength of repulsive interaction. There are two types within this phase. Type I is ``site'' nematic with order parameter, $\eta=\frac{1}{8}\sum_{\delta^\prime}(\langle c^\dagger_{Bi}c_{Bi}\rangle-\langle c^\dagger_{Ci+\delta^\prime}c_{Ci+\delta^\prime}\rangle)$, where $\delta^\prime=\pm \hat{x}/2\pm \hat{y}/2$ denoting four NNN bonds of B-site and $A,B,C$ are sublattice indices.
Without loss of generality, at 2/3 filling we can set the charge density, $\langle c^\dagger_{Ai}c_{Ai}\rangle=\frac{2}{3}+\rho$, $\langle c^\dagger_{Bi}c_{Bi}\rangle=\frac{2}{3}-\frac{\rho}{2}+\eta$, and $\langle c^\dagger_{Ci}c_{Ci}\rangle=\frac{2}{3}-\frac{\rho}{2}-\eta$.
As $\varepsilon_A>0$, at 1/3 filling we simply replace 2/3 by 1/3 in the above expression. Note that the use of the parameter $\rho$ is to take into account the renormalization of the onsite potentials due to interactions. The nonzero expectation value of it does not break any symmetry of the model.
On the other hand, type II is ``bond'' nematic with order parameter either in the form of $Q_1=\frac{1}{4}\text{Re}[\sum_{\delta=\pm\hat{x}}\langle c^\dagger_{Ai}c_{Bi+\delta}\rangle-\sum_{\delta=\pm\hat{y}}\langle c^\dagger_{Ai}c_{Ci+\delta}\rangle]$, or in the form of $Q_2=\frac{1}{4}\sum_{\delta^\prime}D_{\delta^\prime}\text{Re}\langle c^\dagger_{Bi}c_{Ci+\delta^\prime}\rangle$ for $D_{\delta^\prime=\pm(\hat{x}/2-\hat{y}/2)}=1$ and $D_{\delta^\prime=\pm(\hat{x}/2+\hat{y}/2)}=-1$. The subscript of the order parameters indicates their origin of either $V_1$ or $V_2$ repulsion.

\begin{figure}[tbh]
\begin{center}
\includegraphics[scale=0.35]{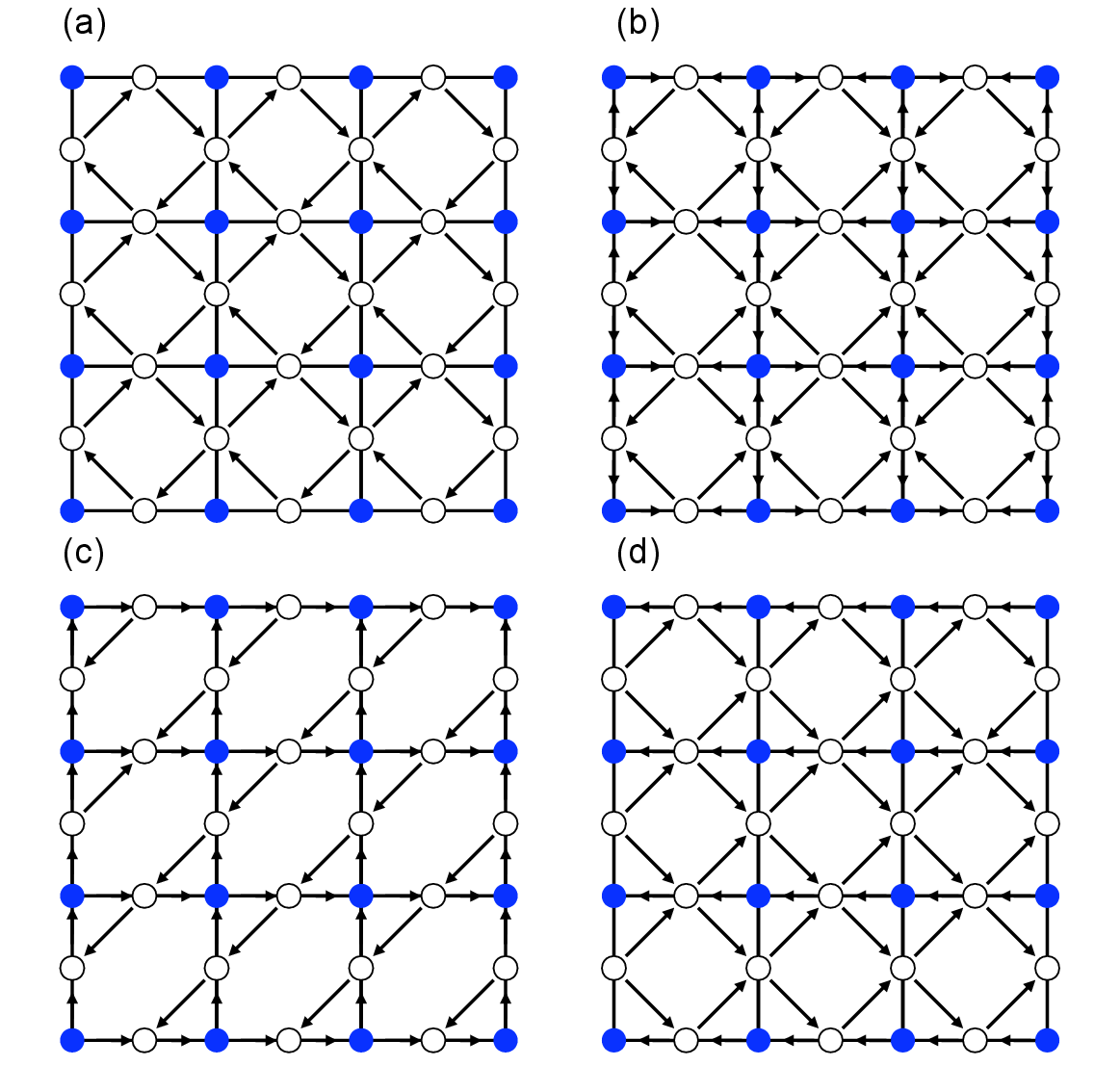}
\caption{(Color online) The current loop ground states for (a) $\Phi_2\neq 0$ and others are zero (QAH state); (b) $\Phi_1,\Phi_2^\prime\neq 0$ and others are zero (Varma $\Theta_I$ loop)(c) $\Phi_1^\prime,\Phi_2^{\prime\prime}=\Phi_2^{\prime\prime\prime}\neq 0$ (Varma $\Theta_{II}$ loop); (d) $\Phi_1^{\prime}=\Phi_1^{\prime\prime},\Phi_2^{\prime\prime\prime}\neq 0$.}
\label{fig:current}
\end{center}
\end{figure}

(ii) {\it Current loop state.} This type of phase is featured by spontaneously TRS breaking. The most probable current patterns which preserve translational invariance with no (charge) source and drain present on the lattice sites are shown in Figs.~\ref{fig:current}(a)-(d). Each state basically comes from non-vanishing imaginary part of certain bond orders in the MF decouplings and may behave Hall-insulating [3(a)], semi-metallic [3(b),3(d)], or insulating [3(c)]. In particular, the most significant one is case (a), which exhibits quantum anomalous Hall (QAH) effect with order parameter,  $\Phi_2=\frac{1}{4}\text{Im}[\sum_{\delta^\prime}D_{\delta^\prime}\langle c^\dagger_{Bi}c_{Ci+\delta^\prime}\rangle]$. This {\it topological} state is known to be characterized by quantized Hall conductance without Landau levels (or equivalently, by non-zero Chern number) and has topology-protected, gapless chiral edge modes.\cite{haldane88} We compute the Chern number for each band within this state and find that 1) for $|\varepsilon_A|>0$, the previous two touching bands now carry Chern numbers $\pm1$ separately. In particular, one of two bands is (nearly) dispersionless. The third one simply carries zero; 2) for $\varepsilon_A=0$, the middle flat band carries zero Chern number, while the upper and lower bands carry $\pm 1$, respectively.

\begin{figure}[t]
\includegraphics[scale=0.35]{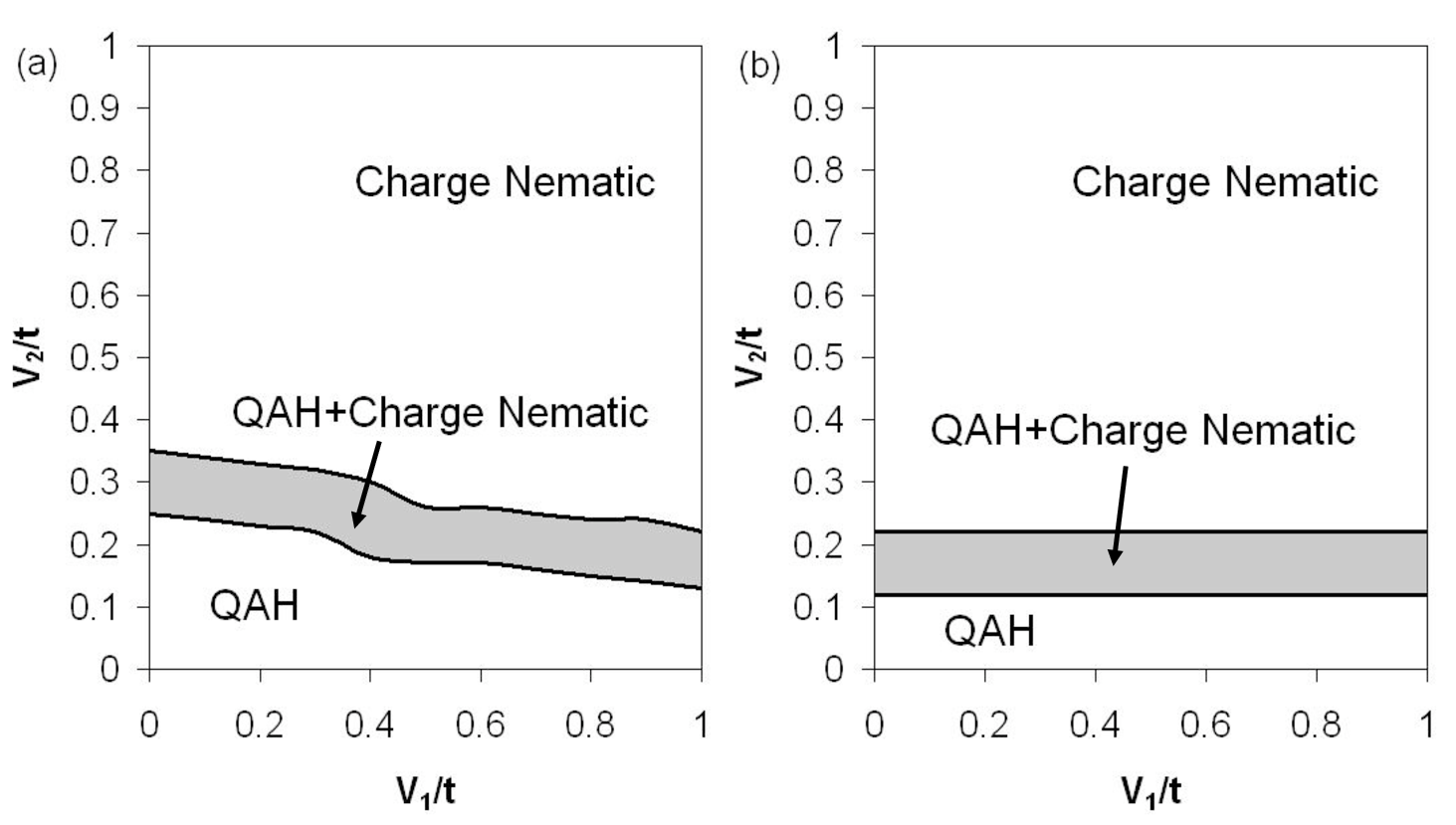}
\caption{(Color online) Schematic zero-temperature mean-field phase diagrams for spinless fermions on the (a) 1/3-filled lattice with $\varepsilon_A/t=4$, $t^{\prime\prime}/t=-0.1$ and on the (b) 2/3-filled lattice with $\varepsilon_A/t=-4$, $t^{\prime\prime}/t=0.1$. The shaded area represents the coexistence region.}
\label{fig:spinlessPD}
\end{figure}

The other possible current loop states, however, are not topological insulating. For case (b) (Varma $\Theta_I$ loop state\cite{varma06}), it is semi-metallic with order parameter given by, simultaneously, $\Phi_1=\frac{1}{4}\text{Im}[\sum_{\delta=\pm\hat{x}/2}\langle c^\dagger_{Ai}c_{Bi+\delta}\rangle-\sum_{\delta=\pm\hat{y}/2}\langle c^\dagger_{Ai}c_{Ci+\delta}\rangle]$ and $\Phi_2^\prime=\frac{1}{4}\sum_{\delta^\prime}\text{Im}\langle c^\dagger_{Bi}c_{Ci+\delta^\prime}\rangle$. For case (c) (Varma $\Theta_{II}$ loop state\cite{varma06}), it is insulating with broken inversion symmetry (IS) as well but is invariant under the combined TRS and IS. Thus, there is no Hall or uniform Kerr response by noticing that for any given momentum $\mathbf{k}$ it changes sign under TRS or IS.\cite{sun08} 
This order can be described as, $\Phi^{\prime}_1\neq 0, \Phi^{\prime\prime}_2=\Phi^{\prime\prime\prime}_2\neq 0$, where $\Phi^\prime_1=\frac{1}{4}\text{Im}[\sum_{\delta=\pm\hat{x}/2}(2\delta\cdot\hat{x})\langle c^\dagger_{Ai}c_{Bi+\delta}\rangle+\sum_{\delta=\pm\hat{y}/2}(2\delta\cdot\hat{y})\langle c^\dagger_{Ai}c_{Ci+\delta}\rangle]$,
$\Phi_2^{\prime\prime}=\frac{1}{4}\text{Im}[\sum_{\delta^\prime}(2\delta^\prime\cdot\hat{x})
D_{\delta^\prime}\langle c^\dagger_{Bi}c_{Ci+\delta^\prime}\rangle]$, and $\Phi_2^{\prime\prime\prime}=\frac{1}{4}\text{Im}[\sum_{\delta^\prime}
(2\delta^\prime\cdot\hat{x})\langle c^\dagger_{Bi}c_{Ci+\delta^\prime}\rangle]$. Finally, case (d) is a semi-metallic state with broken IS and hence no net Hall current in it. Its order is described as $\Phi^{\prime\prime\prime}_2\neq 0, \Phi_1^{\prime}=\Phi_1^{\prime\prime}\neq 0$, where $\Phi^{\prime\prime}_1=\frac{1}{4}\text{Im}[\sum_{\delta=\pm\hat{x}/2}(2\delta\cdot\hat{x})\langle c^\dagger_{Ai}c_{Bi+\delta}\rangle-\sum_{\delta=\pm\hat{y}/2}(2\delta\cdot\hat{y})\langle c^\dagger_{Ai}c_{Ci+\delta}\rangle]$.

\begin{figure}[tbh]
\includegraphics[scale=0.3]{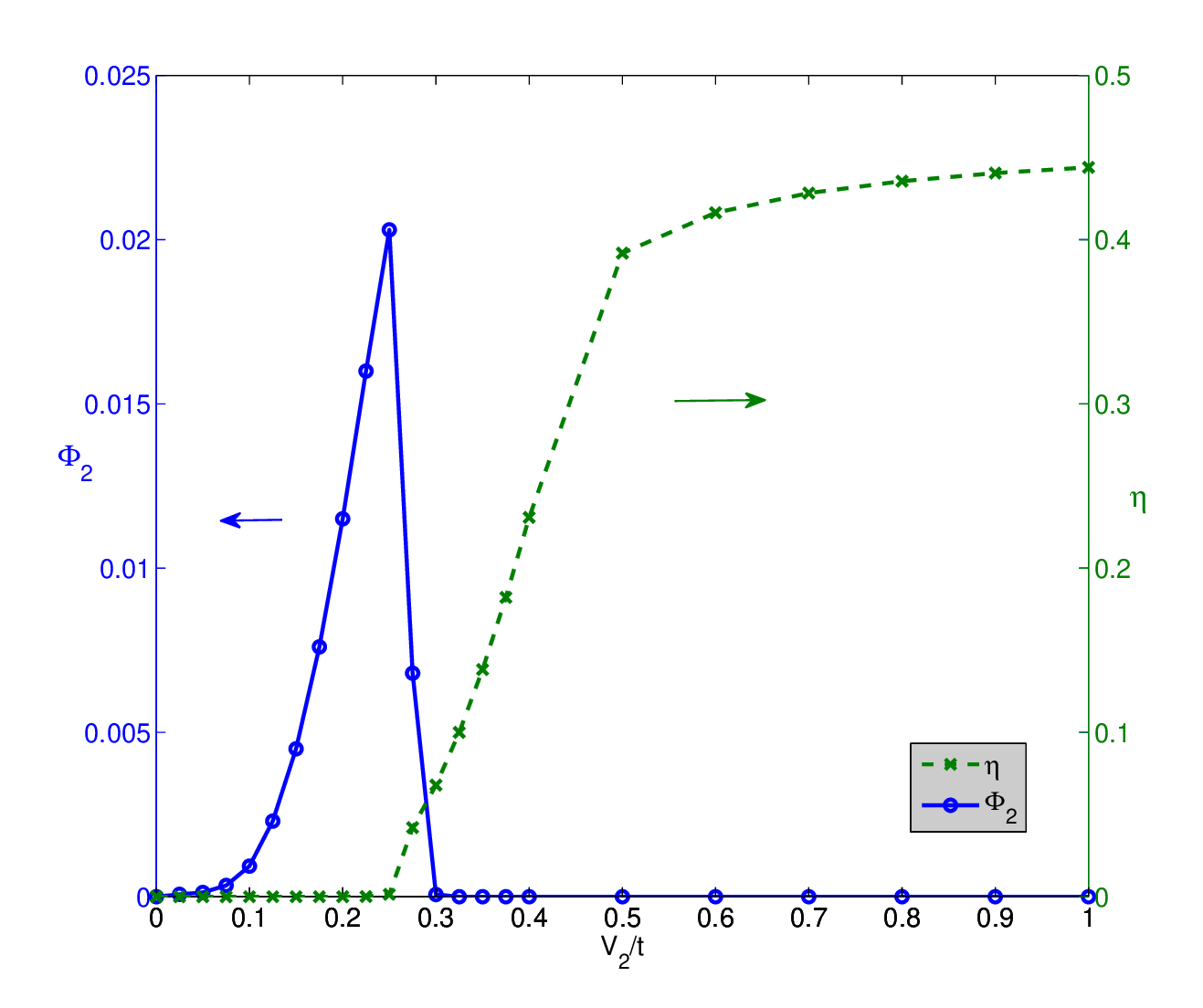}
\caption{(Color online) The magnitude of both QAH (blue, solid) and nematic (green, dashed) order parameters as a function of $V_2$ at 1/3-filled lattice with $\varepsilon_A/t=4$, $t^{\prime\prime}/t=-0.1$, and fixed $V_1/t=0.2$.}
\label{fig:OPvsV2}
\end{figure}

In momentum space, the mean-field Hamiltonian at 2/3 filling can be now written in the matrix form,
\be
\Cal{H}_{MF}=\sum_{\bk}\Psi^\dagger_{\bk}(H_{\bk}-\mu I)\Psi_{\bk}+E_0,
\ee
with the fermion spinor, $\Psi^\dagger_{\bk}=(c^\dagger_{A\bk},c^\dagger_{B\bk},c^\dagger_{C\bk})$,
and
\ba
E_0 &=&-\frac{N\bar{\varepsilon}_{A1}(\bar{\varepsilon}_{B1}+\bar{\varepsilon}_{C1})}{4V_1}
-\frac{N\bar{\varepsilon}_{B2}\bar{\varepsilon}_{C2}}{4V_2} \nonumber \\
&+& 4NV_2(\delta t^{\prime 2} +Q_2^2+\Phi_2^2+\Phi_2^{\prime 2}+\Phi_2^{\prime\prime 2}+\Phi_2^{\prime\prime\prime 2}) \nonumber \\
&+& 4NV_1(\delta t^2+Q_1^2+\Phi_1^2+\Phi^{\prime 2}_1+\Phi^{\prime\prime 2}_1),
\ea
where $\delta t$ ($\delta t^\prime$) represents a renormalization due to NN(NNN) repulsions. The Hamiltonian matrix $H_{\bk}$ reads
\be
H_{\bk}=\left(
\begin{array}
[c]{ccc}
\bar{\varepsilon}_{A1}+\varepsilon_A & \Gamma_x  & \Gamma_y \\
 & \bar{\varepsilon}_{B1}+\bar{\varepsilon}_{B2}+\nu_y & \Gamma_{xy} \\
 &  & \bar{\varepsilon}_{C1}+\bar{\varepsilon}_{C2}+\nu_x,
\end{array}
\right),
\ee
where $\bar{\varepsilon}_{A1}=V_1(8-6\rho)/3$, $\bar{\varepsilon}_{B1}=2V_1(2+3\rho)/3$,
$\bar{\varepsilon}_{B2}=2V_2(4-3\rho-6\eta)/3$,
$\bar{\varepsilon}_{C1}=2V_1(2+3\rho)/3$, $\bar{\varepsilon}_{C2}=2V_2(4-3\rho+6\eta)/3$, and $\nu_{x,y}=-2t^{\prime\prime}\cos k_{x,y}$; $\Gamma$
parameters are given by
\ba
\Gamma_{x,y} &=&
(-2t\mp2V_1(Q_1-i\Phi_1))\cos\frac{k_{x,y}}{2}\nonumber \\
&-& 2V_1(\Phi^\prime_1\pm\Phi^{\prime\prime}_1)\sin\frac{k_{x,y}}{2},
\nonumber \\
\Gamma_{xy} &=& 4iV_2\Phi_2^\prime\cos\frac{k_x}{2}\cos\frac{k_y}{2}
+4V_2\Phi_2^{\prime\prime\prime}\sin\frac{k_x}{2}\cos\frac{k_y}{2}
\nonumber \\
&-& 4V_2(Q_2-i\Phi_2)\sin\frac{k_x}{2}\sin\frac{k_y}{2}\nonumber \\
&-& 4V_2\Phi_2^{\prime\prime}\cos\frac{k_x}{2}\sin\frac{k_y}{2}.
\ea
Thus, the mean-field free energy can be expressed as
\be
F=-\frac{1}{\beta N}\sum_{\bk}\ln (1+e^{-\beta (E_{\bk}-\mu)})+E_0,
\label{eq:free_energy}
\ee
where $\beta=1/k_B T$ and $E_{\bk}$ are eigenvalues of $H_{\bk}$. The ground state with given coupling parameters can then be determined by minimizing the free energy with respect to each order parameter, yielding a set of coupled gap equations. Notice that at 1/3 filling we follow the same procedure and only the diagonal part of $H_{\bk}$ and $E_0$ need to be changed accordingly due to the shift of the average charge density.

We numerically solve the coupled gap equations self-consistently and obtain the zero-temperature $V_1$-$V_2$ phase diagrams at both 1/3 ($\varepsilon_A=4t$) and 2/3 ($\varepsilon_A=-4t$) fillings, as seen in Figs.~\ref{fig:spinlessPD}(a) and (b). Note that in this study, only weak short-range repulsions, i.e., $V_1,V_2\le t$ are considered. At 1/3 filling, we find that there are three phases in the absence of $V_1$: QAH phase, coexisting QAH+nematic phase, and nematic phase. Beginning with $V_2\ll t$, the infinitesimal instability of QBCP leads to QAH phase by the second order phase transition, with a $T=0$ gap, $\Delta_{QAH}=(V_2\Phi_2)\propto\Lambda\text{exp}(\frac{-1}{N_0 V_2})$, where $N_0$ denotes the finite DOS at QBCP and $\Lambda$ is an energy cutoff; on the other hand, for $t\gtrsim V_2 > V_{2c}\sim 0.22t$ the ground state breaks $C_4$ symmetry spontaneously down to $C_2$ and exhibits insulating nematic phase with a gap $\Delta_{N}\sim \eta$. In this phase, we find that the site-nematic order ($\eta$) is the dominant one and a small component of the bond-nematic order ($Q_1$) accompanies with it. In fact, we notice that the bond-nematic order cannot be induced by $V_1$ itself.

Finally, with intermediate value of $V_2$, there exists a narrow window for the coexistence of both QAH and nematic orders. This can be further seen in Fig.~\ref{fig:OPvsV2}, showing the magnitude of both order parameters as a function of $V_2$ with fixed $V_1$. Since the bulk energy gap never really closes as $V_2$ increases, it suggests that the quantum phase transition between QAH and nematic phases is {\it not} continuous, lacking a quantum critical point. In addition, we also notice that there is no room for the current loop states other than QAH state for systems with relatively large $|\varepsilon_A|$ and weak repulsions. 

At 2/3 filling, the phase diagram is qualitatively similar to that at 1/3 filling. However, there are a few remarks worth mentioning here: 1) Although the non-interacting energy spectrum for both fillings can be related by translating ``particle'' into ``hole'' language, which causes $t\rightarrow -t$, $t^{\prime\prime}\rightarrow -t^{\prime\prime}$, $\varepsilon_A\rightarrow -\varepsilon_A$, and $\mu\rightarrow -\mu$. The interactions given in the present form ruin such relation and hence the two phase diagrams must be different.\cite{particlehole}
2) Notice that since in our consideration $|\varepsilon_A|$ is the largest energy scale among others, it is easy to realize that the charge density at A-site is $\delta n$ at 1/3 filling and $1-\delta n$ at 2/3 filling, where $\delta n$ denotes small density fluctuation. Such fact makes the NN repulsion ($V_1$ term) almost a constant depending on the total number of fermions, leading to $V_1$-insignificant phase diagrams in both cases. However, a close study in energetics (assuming the system is in QAH phase) can show that $V_1$ enters the dynamics through the first order of $\delta n$ for 1/3-filling while through the second order of $\delta n$ for 2/3-filling. Therefore, the phase diagram is relatively insensitive to $V_1$ in the 2/3-filling case.
3) When $\varepsilon_A,t^{\prime\prime}\rightarrow 0$, the spin-1 structure near band touching point, as discussed in the previous section, is recovered. Our mean-field study shows that the infinitesimal instability (near 1/3 or 2/3 filling) is absent due to the vanishing DOS of the dispersive bands and the semi-metallic phase is robust until $V_2$ reaches certain critical value. Moreover, for $V_2\geq V_{2c}$, we find that the QAH phase only survives in a negligible window of $V_2$, and the nematic phase becomes the dominant one in the phase diagram (not shown). This result is similar to the work done by Q. Liu {\it et al.}\cite{liu10} on the 2/3-filled kagome lattice with Dirac BCPs.

\subsection{Spinful fermions}
We now take the spin degrees of freedom into account. The model Hamiltonian again consists of the free and interacting parts, i.e, $\Cal{H}=\Cal{H}_0+\Cal{H}_{int}$. The free part is again given by Eq.~(\ref{eq:H0}) with extra spin index $\sigma$ in the fermion creation/annihilation operators; the interacting terms now contain
\ba
\Cal{H}_{int} &=& \sum_{\langle ij\rangle}\sum_{\sigma,\sigma^\prime}V_{1}n_{i,\sigma}n_{j,\sigma^\prime}+
\sum_{\langle\langle ij\rangle\rangle}\sum_{\sigma,\sigma^\prime}V_{2}n_{i,\sigma}n_{j,\sigma^\prime}
 \nonumber \\
&+& \sum_{i}U_{i}n_{i,\uparrow}n_{i,\downarrow}.
\ea
In addition to $V_1$ ($V_2$) denoting the coupling constant of NN (NNN) repulsion, we also consider the repulsive Hubbard ($U_i$) terms. For simplicity, we will assume uniform onsite repulsions, i.e., $U_i=U$.

Different from the spinless case, there are not only spin-{\it singlet} order parameters (as we had before), but also spin-{\it triplet} order parameters within mean-field approximation. The possible phases under {\it translation-invariant} ansatz are classified below:\cite{FM}

(i) {\it Charge nematic state} (CN). This phase is associated with spontaneously rotational ($C_4$) symmetry breaking. One can either have the site-nematic
or bond-nematic state, whose order parameter is a spin-singlet and simply the same as that for spinless fermions with additional summation on spin $\sigma$ times a normalization factor 1/2. Note that for the site-nematic case, the driving force is now from both $U$ and $V_2$ terms, combined together to give out an effective NNN repulsion, $V_2^\prime=2(V_2-\frac{U}{8})$, playing similar role of $V_2$ in the spinless model.

\begin{figure}[tbh]
\includegraphics[scale=0.4]{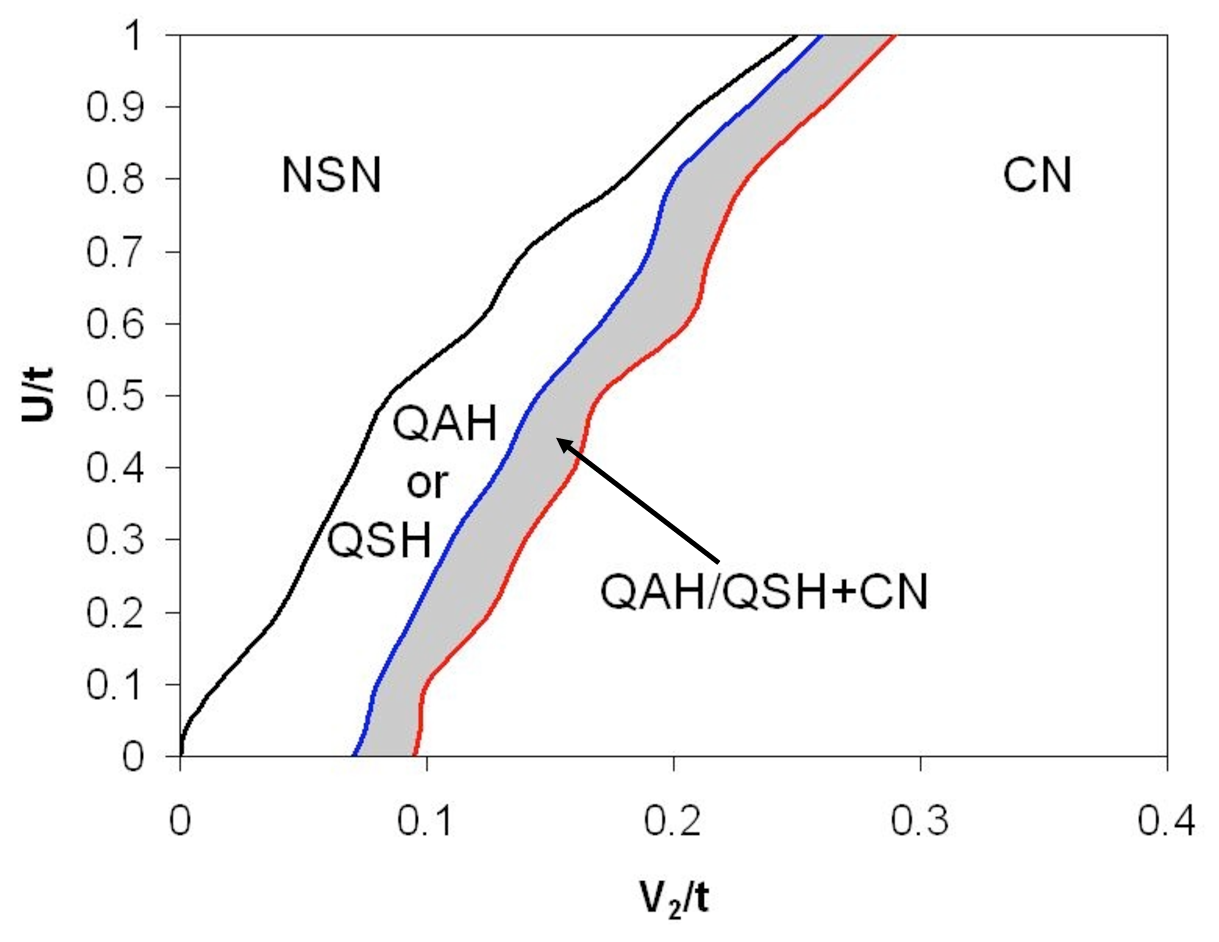}
\caption{(Color online) Schematic zero-temperature mean-field phase diagram for spinful fermions at 2/3-filling with $\varepsilon_A/t=-4$, $t^{\prime\prime}/t=0.1$, and $V_1=0$. The short-hand notations represent: NSN = nematic-spin-nematic; QAH = quantum anomalous Hall; QAH = quantum anomalous Hall; CN = charge nematic. The shaded area denotes the coexistence region.}
\label{fig:spinfulPD}
\end{figure}

(ii) {\it Nematic-spin-nematic state} (NSN).
This phase breaks $C_4$ symmetry in the spin sector, {\it not} in the charge sector. Consequently, it turns the (spin) doubly degenerate QBCP into four Dirac points (two pairs with opposite spin polarizations), and $C_4$ symmetry of the band structure remains intact.
Similar to its charge counterpart in (i), there are two types: One is the site-NSN with a spin-triplet order parameter, $\vec{\eta}^t$=$\frac{1}{16}\sum_{\delta^\prime}
(\mathbf{S}_{Bi}-\mathbf{S}_{Ci+\delta^\prime})$, where $\mathbf{S}_\alpha$=$\langle c^\dagger_{\alpha i,\sigma}\mathbf{s}_{\sigma\sigma^\prime}c_{\alpha i,\sigma^\prime}\rangle$ with $\alpha=A,B,C$ and $\mathbf{s}$, the Pauli matrices. Note that this phase can occur simply due to the presence of the Hubbard term, which provides a spin-triplet channel, $-\frac{2}{3}U(\vec{S}_i)^2$ with $\vec{S}_i$ representing usual spin operator. The other one is the bond-NSN, which is described by $\vec{Q}^t_1$=$\frac{1}{8}\text{Re}$[
$\sum_{\delta=\pm\hat{x}/2}\langle c^\dagger_{Ai,\sigma}\mathbf{s}_{\sigma\sigma^\prime}
c_{Bi+\delta,\sigma^\prime}\rangle$-$\sum_{\delta=\pm\hat{y}/2}\langle c^\dagger_{Ai,\sigma}\mathbf{s}_{\sigma\sigma^\prime}
c_{Ci+\delta,\sigma^\prime}\rangle]$ or $\vec{Q}^t_2$=$\frac{1}{8}\sum_{\delta^\prime}D_{\delta^\prime}\text{Re}\langle c^\dagger_{Bi,\sigma}\mathbf{s}_{\sigma\sigma^\prime}c_{Ci+\delta,\sigma^\prime}
\rangle$ (breaking $C_4$ along a diagonal direction of the lattice).

(iii) {\it Charge current-loop state with broken TRS}.
As mentioned in the spinless model, among all the current patterns the case (a) in Fig.~\ref{fig:current} is of most interest. This state, characterized by spontaneously broken TRS and parity symmetry, exhibits QAH effect with a spin-singlet order parameter,  $\Phi_2=\frac{1}{8}\text{Im}[\sum_{\delta^\prime,\sigma}D_{\delta^\prime}\langle c^\dagger_{Bi,\sigma}c_{Ci+\delta^\prime,\sigma}\rangle]$. Mainly driven by $V_2$ terms, fermions with opposite spin polarizations flow in the same way and hence provide the same flux pattern, penetrating the whole lattice. In addition, we investigate the possibility for the other current-loop states [e.g., from Fig.~\ref{fig:current}(b)-(d)] and find that none of them is stabilized by the presence of short-range repulsions in our MF study. Therefore, we will only consider case (a) hereafter.

(iv) {\it Spin current-loop state with TRS}.
The key difference of this phase from its charge counterpart (QAH) is TRS unbroken. One can view it as a combined double-layer QAH system: Fermions of opposite spin polarizations, residing in different layers, producing just opposite flux patterns separately. Thus, for the whole system TRS is preserved. In fact, this is known as quantum spin Hall (QSH) phase, or equivalently, 2D Z$_2$-nontrivial TI, with spin-triplet order parameter described by $\vec{\Phi}^t_2=\frac{1}{8}\sum_{\delta^\prime}D_{\delta^\prime}\text{Im}[\langle c^\dagger_{Bi,\sigma}\mathbf{s}_{\sigma\sigma^\prime}
c_{Ci+\delta^\prime,\sigma^\prime}\rangle]$. Two remarks deserve mentioning here. First, both QAH and QSH are {\it topological} phases, characterized by non-trivial topological index with robust edge states. However, the former one acquires non-vanishing Chern number, while the latter one has zero Chern number due to TRS. Thus, a new topological index, called Z$_2$ index $\nu$, needs to be introduced.\cite{kane05a,kane05b,fu07b} As detailed in Appendix B, the QSH phase on the Lieb lattice indeed acquires non-trivial $\nu=1$. Second, it is straightforward to see that at MF level, the energy spectra (not shown here) for both QAH and QSH are the same. As a result, they have equal energy gain from $V_2$ repulsion and hence one cannot distinguish them in the MF phase diagram. If there were an extra NNN exchange coupling $J_2$ present in the system, the QAH would be favored for $J_2>0$; reversely, the QSH would be favored for $J_2<0$ due to its spin-triplet nature.

Under our assumption of translational invariance within our mean-field study, we do not consider any charge or spin density wave order. However, it is still worth pointing out that if $|\varepsilon_A|\rightarrow\infty$ and hence makes the $A$ sublattice be effectively decoupled from rest of the lattice sites, at large $U$ the (0,0) antiferromagnetic order could be realized  at 1/3 (2/3) filling with $\varepsilon_A>0$ ($\varepsilon_A<0$), as guaranteed by the Lieb's theorem.\cite{lieb89}


Following the same procedure as in the spinless case, we decouple the interacting terms within MF approximation and obtain the MF free energy in similar form of Eq.~(\ref{eq:free_energy}), where we have used the MF ansatz, $\langle n_{Ai,\sigma}\rangle=\frac{2}{3}+\rho+\frac{1}{2}\sigma S_{Az}$, $\langle n_{Bi,\sigma}\rangle=\frac{2}{3}-\frac{\rho}{2}+\eta+\frac{1}{2}\sigma (S_{Bz}+2\eta^t_z)$, and $\langle n_{Ci,\sigma}\rangle=\frac{2}{3}-\frac{\rho}{2}-\eta+\frac{1}{2}\sigma (S_{Bz}-2\eta^t_z)$ for the 2/3-filled lattice. Note that, for simplicity, we have assumed that spin points to $z$-direction after SU(2) symmetry breaking. Such treatment will be applied to other spin-triplet order parameters as well. $E_0$ in the spinful case becomes
\ba
\frac{E_0}{N} &=&-\frac{\bar{\varepsilon}_{A1}(\bar{\varepsilon}_{B1}+\bar{\varepsilon}_{C1})}{V_1}
-\frac{\bar{\varepsilon}_{B2}\bar{\varepsilon}_{C2}}{V_2}+8V_1(\delta t^2+Q_1^2+Q_{1z}^{t2}) \nonumber \\
&+& 8V_2(\delta t^{\prime 2} +Q_2^2+\Phi_2^2+Q_{2z}^{t2}+\Phi_{2z}^{t2}) \nonumber \\
&+& \frac{U}{4}(S^2_{Az}+S^2_{Bz}+S^2_{Cz}+8\eta^{t2}_{z}) \nonumber \\
&-& U[\frac{(\bar{\varepsilon}_{B1}+\bar{\varepsilon}_{C1})^2}{16V_1^2}
+\frac{\bar{\varepsilon}_{B2}^2+\bar{\varepsilon}_{C2}^2}{16V_2^2}],
\ea
where $\bar{\varepsilon}_{A1}, \bar{\varepsilon}_{B1},\bar{\varepsilon}_{B2},
\bar{\varepsilon}_{C1}$, and $\bar{\varepsilon}_{C2}$ are defined as before.
$E_{\bk}$ are eigenvalues of the MF Hamiltonian, which is now a $6\times 6$ matrix.

By minimizing free energy with respect to various order parameters, the $T=0$ mean-field $U$-$V_2$ phase diagram at 2/3 filling ($\varepsilon_A=-4t$) for spinful fermions is shown in Fig.~\ref{fig:spinfulPD}. We first notice that in the absence of $U$ there again exist three phases: QAH/QSH phase, coexisting QAH/QSH+CN phase, and CN phase from weak to strong $V_2$ repulsion. In particular, the fact that topological QAH/QSH phase can arise from infinitesimal instability of QBCP further justifies the interaction-driven scenario as a promising way for producing TI. In the presence of $U$, however, NSN phase begins to compete with the topological phase and clearly dominates over QAH/QSH whenever $U\gg V_2>0$. On the other hand, as $V_2^\prime=2(V_2-\frac{U}{8})\gtrsim 0.22t$, the insulating CN phase takes over the phase diagram and this is consistent with the result shown in the spinless model. Two remarks are worth mentioning here. The first one is about the effect of NN repulsion $V_1$. As in the spinless case, at 2/3 filling the phase diagram is not sensitive to the presence of $V_1$. Especially, $V_1$ itself does not lead to any order. However, when $V_1$ becomes stronger, it is quite possible that the system might gain certain energy by opening a gap due to translation symmetry breaking, and hence beyond our current consideration. The second point is about the system at 1/3 filling with $\varepsilon_A=4t$. In fact, the phase diagram is qualitatively similar to that of 2/3 filling and therefore we omit it without further discussion.

\section{Discussion and conclusion}
The model we have solved on the Lieb lattice demonstrates that the TI (QAH/QSH state) can be induced by appropriate interactions through spontaneously symmetry breaking mechanism, which dynamically generates spin-orbit couplings necessary for a topological insulator. It is then natural to ask how it can be realized experimentally.

We notice that in the interaction-driven scenario there are two key conditions that a system has to fulfill: 1) The band structure should contain suitable band crossing point at which two touching bands have opposite curvatures. 2) The system should have weak (or no) spin-orbit coupling [spin SU(2) symmetry is preserved] and NNN repulsions need to be more significant than the other short-range repulsions. While the condition 2) is tricky and we have to reserve it for future investigation, we would like to comment on some possible routes for condition 1) below.

\begin{figure}[tbh]
\def\subfigcapskip{0pt}
\def\subfigtopskip{5pt}
\def\subfigbottomskip{5pt}
\subfigure[]{\includegraphics[scale=0.35]{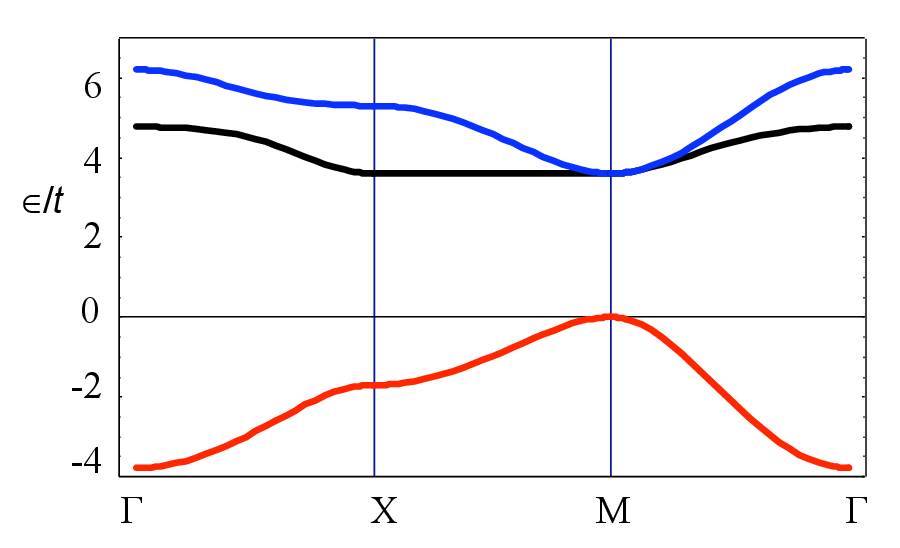}}
\subfigure[]{\includegraphics[scale=0.35]{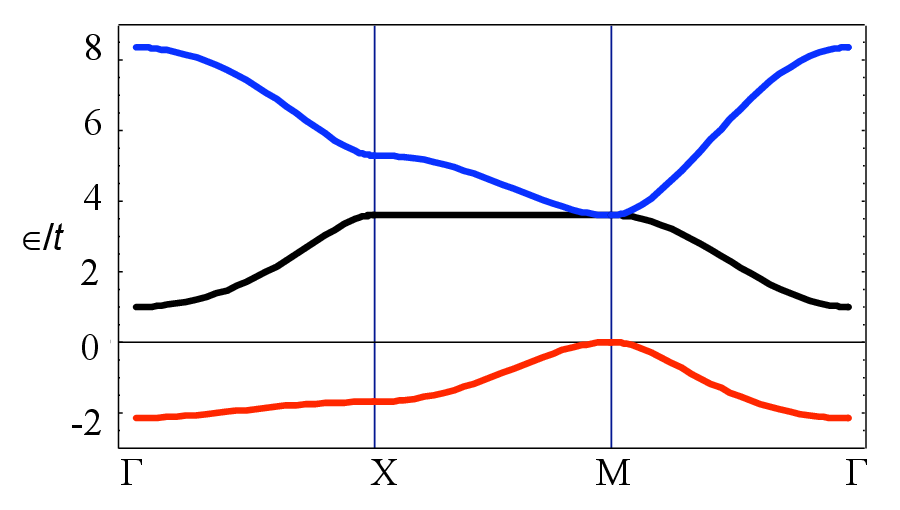}}
\caption{(Color online) (a) Energy spectrum of the model for cuprates along the high-symmetry lines in the BZ with parameters mentioned in the text. (b) The same model as used in (a), but with reversed sign of $t^\prime$.}
\label{fig:spectrumReal}
\end{figure}

First of all, the most promising candidate, we believe, is from cold atom system. As discussed by Goldman et al. in Ref.~\onlinecite{goldman11}, the Lieb lattice may be constructed as an optical lattice created by properly arranged laser beams. In particular, the spin-orbit interaction, which might be an issue in traditional materials, now becomes irrelevant. Another potential way for realizing Lieb lattice may come from layered perovskites. A well-known example is the CuO$_2$ plane in high-$T_c$ cuprates such as La$_{1-x}$Sr$_x$CuO$_4$ or YBa$_2$Cu$_3$O$_7$, whose electronic structure might be captured by a three-band model with $p_x,p_y$ and $d_{x^2-y^2}$ orbitals. Here, for illustration purpose we take a typical three-band (Emery) model, written in {\it hole} language, having the same form of Eq.~(\ref{eq:H0}).\cite{emery87,dagotto94} The corresponding model parameters are given by Hybertsen et al.\cite{hybertsen89}: $t = 1.5$, $t^\prime = 0.65$, $\varepsilon_{A} = 0$, and $\varepsilon_{B,C} = 3.6$, where $t^\prime$ denotes NNN hopping. The band structure in the FBZ along high symmetry lines is presented in Fig.~\ref{fig:spectrumReal}(a). As one can see, there is indeed a BCP at $\mathbf{M}$, but with ``wrong'' curvatures for two touching bands. As a result, cuprates may be impractical to produce topological phase as we desire. To overcome this issue, we offer two speculative suggestions. The first way out is to add decorated elements with suitable orbital nature (e.g., $p$-orbital) between atoms on $B$ and $C$ sites that could change the sign of $t^\prime$. This change leads to the band structure shown in Fig.~\ref{fig:spectrumReal}(b) with ``right'' curvatures now. Keeping the same orbital characters, the second way is to search for a new system among perovskite materials, whose model Hamiltonian is similar to that given by Sun et al.'s recent work in Appendix F.\cite{sun10} The key feature of such three-band model is that the dominating transfer integrals are now associated with distance $a$ after hopping. For instance, if, by certain geometric reason, the relevant orbital on $B$ ($C$) sublattice becomes $p_y$ ($p_x$), instead of $p_x$ ($p_y$) as in the cuprates, the above consideration could be plausible.

Before concluding our work, we would like make a brief remark for the possible new physics brought by the (nearly) flat band appearing in the Lieb lattice. Consider $\varepsilon_A>0$ and only NN hoppings in our spinless model. Now, the middle band is completely flat and touches with the lower dispersive band at $\mathbf{M}$ in FBZ. At 1/3 filling if we turn on $V_2$, the ground state would enter QAH state with an energy gap $\Delta_{QAH}$ opened at $\mathbf{M}$. As we have mentioned in Sec. III A, the middle and lower bands acquire Chern numbers $\pm 1$,  while the upper one has zero Chern number in this phase. Moreover, the flatness of the middle band is approximately determined by the ratio of band-gap to bandwidth $\sim\Cal{O}(\epsilon_A/\Delta_{QAH})$. In other words, with appropriately chosen $\epsilon_A/\Delta_{QAH}$, one can produce a nearly flat band with non-trivial Chern number. This result is quite significant since it may provide an opportunity to realize {\it fractional} QAH state (or {\it fractional} Z$_2$ TI in the spinful case) when such band is partially filled.\cite{tang11,neupert11,sun11,sheng11,bernevig11,lu11}

In conclusion, we have studied interacting spinless/spinful fermions on the (extended) Lieb lattice and have explored the possibilities of various spontaneously broken symmetries associated with a BCP in the band structure. Due to the topological nature of the BCP, namely, with Berry phase $2\pi$, we have seen that in the $T=0$ phase diagram the system can exhibit topological QAH/QSH, nematic, and nematic-spin-nematic phases, depending on the strengths of short-range density-density interactions. In particular, the existence of TI phase firmly justifies the interaction-driven scenario.
Moreover, for a quadratic BCP (as $\epsilon_A, t^{\prime\prime}\neq 0$), only weak interaction is necessary for inducing the TI phase, which is in sharp contrast to the systems with Dirac points. In addition, in our model there exists a {nearly} flat band, which is interesting on its own right and we argue that in principle, one can obtain a nearly flat {\it topological} band without external magnetic field, a starting point to realize exotic correlated phases of matter. Still, there are many open issues and they deserve further investigation. For instance, one could consider the effect of the chemical potential away from the BCP or the instability to superconductivity from both repulsions and attractions, with special focus on the possibility of any topological nature. 

\begin{acknowledgments}
We thank L. Fu, H. Lin, C.-K. Lu, Y. Ran, K. Sun, and C. Xu for helpful discussions and, especially, for K. Sun's early collaboration in this work. WFT would like to thank for the hospitality of IoP, Chinese Academy of Sciences during his visit, where part of the present work was done. This work was supported in part by the MOST in Taiwan with Grant No. 103-2112-M-110-008-MY3 (WFT) and by the NSFC under Grant No.11474175 at Tsinghua (HY).

{\it Note added.} After this work is completed, 
we were informed by Steve Kivelson that there is a partly related 
paper by M. H. Fischer and Eun-Ah Kim, arXiv:1106.6060.
\end{acknowledgments}
\appendix
\section{Perturbative renormalization group analysis for a quadratic band crossing point}
In this section, we derive RG equations given in Eq.~(\ref{eq:RGeq}) in path integral formulation and show the RG flows for two typical cases. We begin with defining the action and the shorthand notations below. At zero temperature, the full action is given by
\be
S=S_0+S_{int},
\ee
where the free action $S_0$ reads
\be
S_0=\sum_{\sigma}\int\frac{d\omega}{2\pi}\int\frac{d^2 p}{(2\pi)^2}\bar{\Psi}_{\sigma}(\omega,\mathbf{p})[\hat{G}^{\sigma}_{0}
(\omega,\mathbf{p})]^{-1}
\Psi_{\sigma}(\omega,\mathbf{p}),
\ee
in momentum space. The Grassmann variables, $\bar{\Psi}_\sigma=(\bar{\psi}_{1\sigma},\bar{\psi}_{2\sigma})$ with 1,2 (labeled by `$a$' for later use) representing orbital degrees of freedom. As mentioned in the main text, for simplicity, we consider the inverse of the non-interacting Green's function $[\hat{G}^{\sigma}_{0}
(\omega,\mathbf{p})]^{-1}=i\omega I -d_x\sigma_x-d_z\sigma_z$, which is suitable for a particle-hole symmetric QBCP with $d_x=p_x p_y/m_0$, $d_z=(p_x^2-p_y^2)\bar{t}/(2m_0)$, and $\bar{t}=2$ and is independent of spin polarization $\sigma$. We will fix $1/(2m_0)$ as our unit in the analysis. The action of interactions can be written as,
\ba
S_{int}&=& S_{1}+S_{2},  \\
S_1 &=& -\sum_{a=1}^{2}\int d\xi
u\bar{\psi}_{a\uparrow}(4)\bar{\psi}_{a\downarrow}(3)\psi_{a\downarrow}(2)
\psi_{a\uparrow}(1), \nonumber \\
S_2 &=& -\sum_{\sigma\sigma^\prime}\int d\xi g_{\sigma\sigma^\prime}
\bar{\psi}_{1\sigma}(4)\bar{\psi}_{2\sigma^\prime}(3)\psi_{2\sigma^\prime}(2)
\psi_{1\sigma}(1), \nonumber
\ea
where $\int d\xi=\prod_{b=1}^{3}\int_{-\infty}^{\infty}\frac{d\omega_b}{2\pi}\int\frac{d^2 p_b}{(2\pi)^2}$ and the shorthand notation $b=(\omega_b,\mathbf{p}_b)$. Note that ``4=1+2-3'' is understood by energy and momentum conservation and we 
ignore the momentum dependence of $u$ and $g_{\sigma\sigma^\prime}$, which turns out to be irrelevant in RG sense. In addition, it is worth mentioning here that the present RG analysis is much simpler than that of usual Fermi liquids in the sense that there is {\it no} Fermi surface ($\mathbf{p}_F=0$). This fact thus waives the complexity brought by the Fermi surface, a situation similar to $\phi^4$-theory.

\begin{figure}[bht]
\includegraphics[scale=0.5]{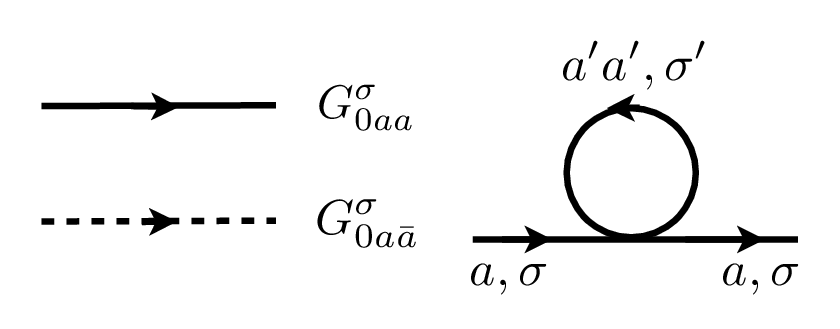}
\caption{The tadpole diagram, induced by the quartic terms, can in principle renormalize the free action. The arrowed lines on the left represent the intra-orbital (solid) and inter-orbital (dashed) non-interacting Green's functions, respectively.}
\label{fig:tadpole}
\end{figure}

Let us now sketch how to obtain one-loop RG equations, following the standard perturbative RG procedure by Shankar.\cite{shankar94} Defining a momentum cutoff $\Lambda$ and $\psi_<$ ($\psi_>$) as slow (fast) modes with $0<p<\Lambda/s$ ($\Lambda/s<p<\Lambda$), the key formula we use to derive the renormalized action by integrating out the fast modes is given by
\ba
Z &=& \int\Cal{D}[\bar{\psi}_>\psi_>;\bar{\psi}_<\psi_<]
e^{S[\bar{\psi}_<+\bar{\psi}_>,\psi_<+\psi_>]} \nonumber \\
&=& \int\Cal{D}[\bar{\psi}_<\psi_<]e^{S^\prime[\bar{\psi}_<\psi_<]},
\ea
where
\ba
e^{S^\prime[\bar{\psi}_<,\psi_<]}&=& Z_0(\Lambda,\Lambda/s)
e^{S_0[\bar{\psi}_<,\psi_<]}\langle e^{S_{int}}\rangle_{0,>} \nonumber \\
&=& Z_0(\Lambda,\Lambda/s)
e^{S_0[\bar{\psi}_<,\psi_<]}  \\
&\times& e^{[\langle S_{int}\rangle_{0,>}+\frac{1}{2}(\langle S_{int}^2\rangle_{0,>}-\langle S_{int}\rangle_{0,>}^2)+\cdots]}. \nonumber
\ea
Note that in deriving last equality we have used the cumulant expansion. $Z_0$ denotes the non-interacting partition function and $\langle\quad\rangle_{0,>}$ represents average with respect to the fast modes of action $S_0$. The collected exponent of the exponential terms now results in the renormalized action $S^\prime$.

Before examining the renormalization of coupling parameters $u$ and $g_{\sigma\sigma^\prime}$, we first notice that, to one-loop level, the quartic terms in $S_{int}$ do induce quadratic terms from the tadpole diagram shown in Fig.~\ref{fig:tadpole}, which are momentum independent. This would indicate there is no non-interacting fixed point in our system. However, explicit calculations show that all contributions from such type of diagram are zero and hence our starting FP survives without flowing ($\bar{t}$, $\psi_{a\sigma}$ are not renormalized), in contrast to the case of the Luttinger liquid.

\begin{figure}[bht]
\includegraphics[scale=0.3]{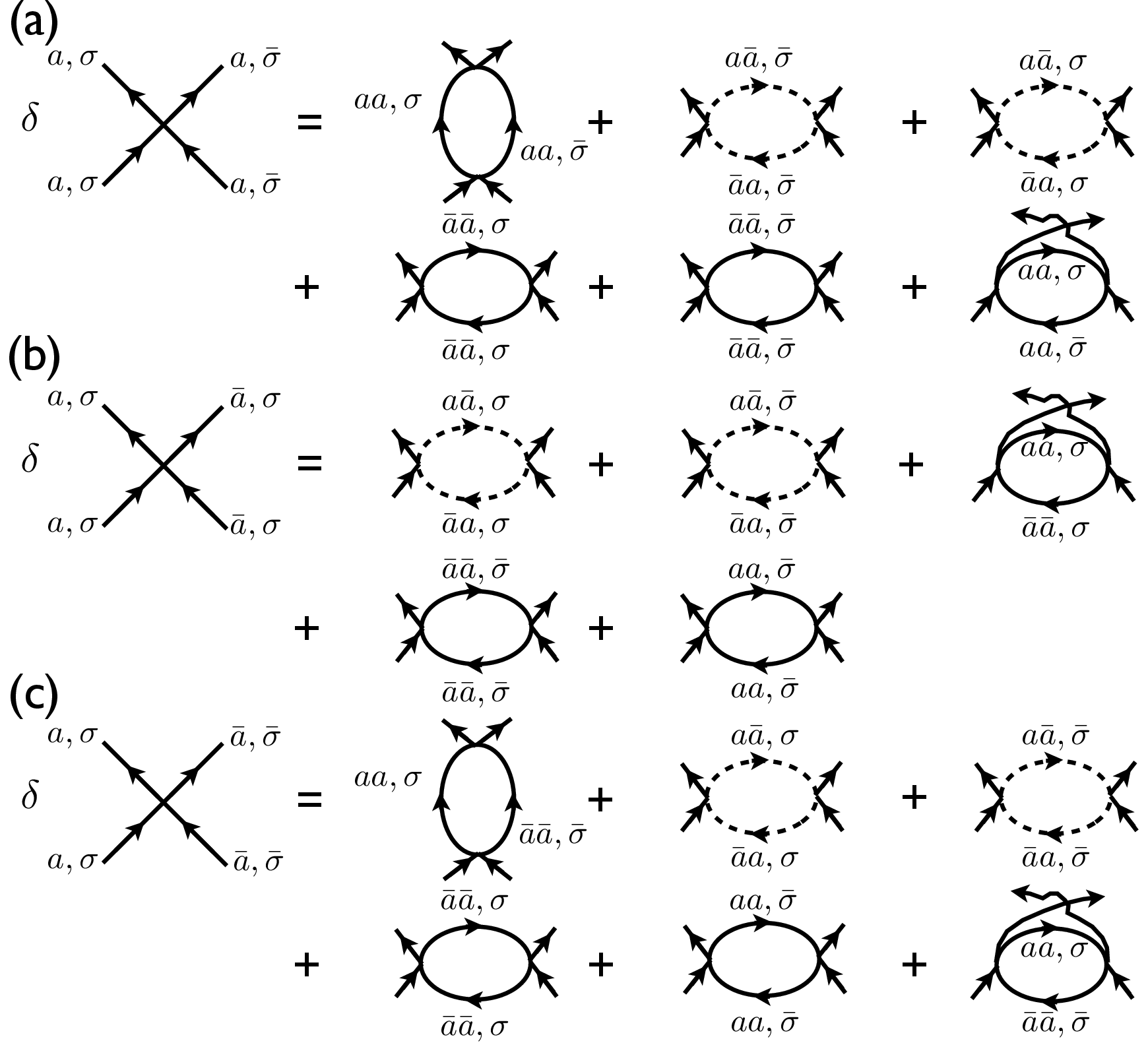}
\caption{Diagrams to one-loop level which contribute to renormalize (a) $u$, (b) $g_{\sigma\sigma}$, and (c) $g_{\sigma\bar{\sigma}}$. Different types of vertices are shown on the left. The labels for external legs in each one-loop diagram are the same as the vertex on the left and hence are omitted.}
\label{fig:diagram}
\end{figure}

Next, we turn our attention to the diagrams that renormalize coupling parameters. As shown in Figs.~\ref{fig:diagram}(a)-(c), they are, respectively, total contributions to the renormalized $u$, $g_{\sigma\sigma}$, and $g_{\sigma\bar{\sigma}}$. Obviously, they have standard structures such as `BCS', `ZS' (zero sound), and `ZS$^\prime$' used in Shankar's seminal work.\cite{shankar94} Note that we have taken the convention that $\bar{1}=2,\bar{\uparrow}=\downarrow$ and vice versa. The essential fermion loop integrals are listed below (without vertex):
\ba
\text{BCS1}&:&
\int_q G_{0aa}^{\sigma}(q)G_{0\bar{a}\bar{a}}^{\sigma^\prime}(-q) =
dl\cdot \gamma(\bar{t}), \nonumber \\
\text{BCS2}&:&
\int_q G_{0a\bar{a}}^{\sigma}(q)G_{0\bar{a}a}^{\sigma^\prime}(-q) =
dl\cdot \gamma(\bar{t}), \nonumber \\
\text{BCS3}&:&
\int_q G_{0aa}^{\sigma}(q)G_{0aa}^{\bar{\sigma}}(-q)  =
dl\cdot[\alpha(\bar{t})-\gamma(\bar{t})], \nonumber \\
\text{ZS1}&:&
\int_q G_{0aa}^{\sigma}(q)G_{0\bar{a}\bar{a}}^{\sigma^\prime}(q) =
-dl\cdot[\alpha(\bar{t})-\gamma(\bar{t})], \nonumber \\
\text{ZS2}&:&
\int_q G_{0a\bar{a}}^{\sigma}(q)G_{0\bar{a}a}^{\sigma}(q) =
dl\cdot\gamma(\bar{t}), \nonumber \\
\text{ZS3} &:&
\int_q G_{0aa}^{\sigma}(q)G_{0aa}^{\sigma^\prime}(q)  =
-dl\cdot\gamma(\bar{t}), \nonumber
\ea
where $\int_q\equiv\int^{\infty}_{-\infty}\frac{d\omega}{2\pi}\int^{2\pi}_{0}
\frac{d\theta}{2\pi}\int_{\Lambda/s}^{\Lambda}\frac{dq}{2\pi}$ and $\bar{t}$, $\alpha(\bar{t})$, $\gamma(\bar{t})$ are defined in the main text [just below Eq.~(\ref{eq:RGeq})]. The superscript, $\sigma^\prime$, of $G_0$ can be either $\sigma$ or $\bar{\sigma}$. The presence of $dl$ is due to the approximation we have used, $\ln s=\ln(1+dl)\approx dl$ in the limit $s\rightarrow 1$. In order to evaluate these integrals, we have set all external $\omega=p=0$ and made $q_x=q\cos\theta,q_y=q\sin\theta$. Combining all the Feynman diagrams in Fig.~\ref{fig:diagram} and keeping in mind the order of fermion operators, the straightforward algebra gives rise a set of RG equations, as we desire, Eq.~(\ref{eq:RGeq}).

\begin{figure}[tbh]
\includegraphics[scale=0.45]{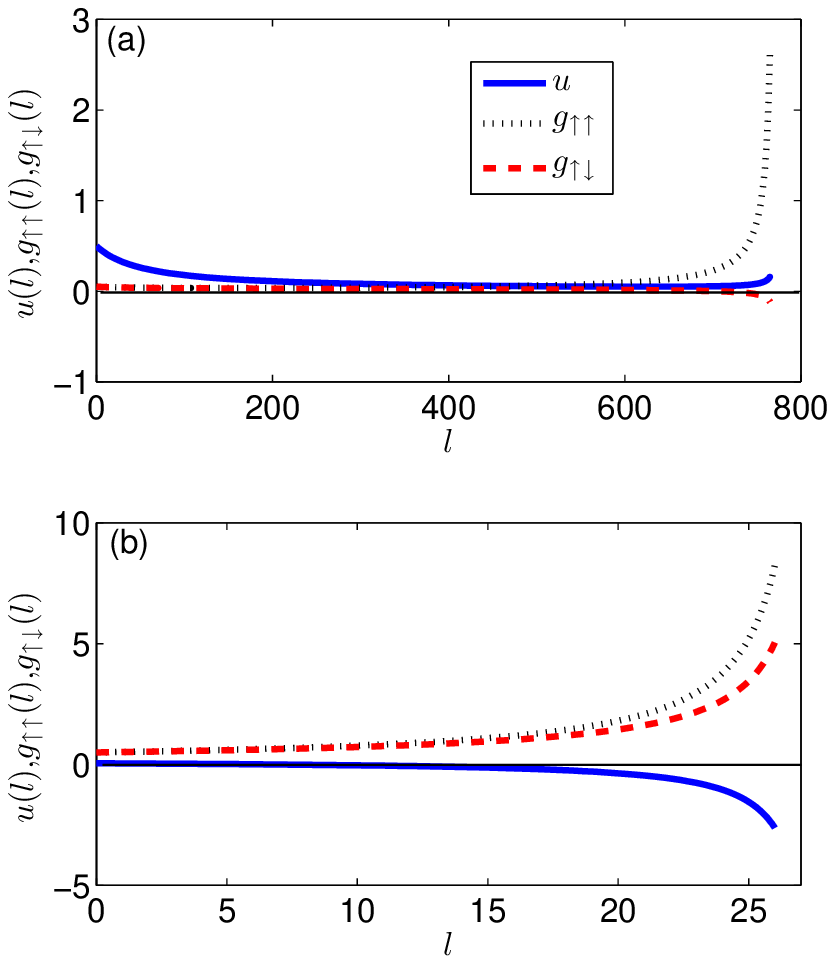}
\caption{(Color online) RG flows for $u(l)$ (solid,blue), $g_{\uparrow\uparrow}(l)$ (dotted,black), and $g_{\uparrow\downarrow}(l)$ (dashed,red) with interaction profile (a) $u(0)=0.5,g_{\uparrow\uparrow}(0)=g_{\uparrow\downarrow}(0)=0.05$;
(b) $u(0)=0.05,g_{\uparrow\uparrow}(0)=g_{\uparrow\downarrow}(0)=0.5$. In both cases, $\bar{t}=2$.}
\label{fig:RGflow}
\end{figure}

Finally, we investigate the RG flow beginning with weak coupling regime for two typical cases: (a) $u>g_{\sigma\sigma}=g_{\sigma\bar{\sigma}}$; (b) $u<g_{\sigma\sigma}=g_{\sigma\bar{\sigma}}$. As one can see in Fig.~\ref{fig:RGflow}, in either case, they all flow to strong coupling regime and hence the non-interacting fixed point is actually unstable against short-range repulsions.

\section{The proof of non-trivial $Z_2$ index for the quantum spin Hall phase}
The existence of the QSH state can be justified by a non-trivial $Z_2$-valued invariant $\nu=1$. This $Z_2$ index represents the topological nature of the QSH state and leads to topologically protected gapless edge states when the 2D system is made open with boundaries. Here, we present a method, first invented by Fu and Kane\cite{fu07b}, to derive $Z_2$ index explicitly. This method essentially takes the advantage of {\it inversion symmetry} such that the index can be related to the parity eigenvalues $\xi_{2m}(\Gamma_i)$ of $2m$ occupied energy bands at four TRIM: $\Gamma_i=\mathbf{\Gamma},\mathbf{X},\mathbf{Y}$, and $\mathbf{M}$.

For the purpose of instruction, we focus on 1/3 filling with $\varepsilon_{A,B,C}=0$ and hence only the lower band is relevant. As shown in Sec III, the presence of $V_2$ can dynamically generate the ``spin-orbit'' interaction, which is equivalent to add the following term into Eq.~(\ref{eq:h0}).
\be
\Cal{H}_{SO}=\sum_{\mathbf{k},\sigma}\psi_{\mathbf{k}\sigma}^\dagger H_{SO}(\mathbf{k})\psi_{\mathbf{k}\sigma} \label{eq:hso}
\ee
with $H_{SO}(\bk)=$
\be
\pm \lambda_{SO}\left(
\begin{array}
[c]{ccc}
0 & 0 & 0\\
0 & 0 & 2iw_{\bk}\\
0 & -2iw_{\bk} & 0
\end{array}
\right),
\ee
where $w_\bk\equiv\sum_{j=1}^2(-1)^{j+1}\cos(\bk\mathtt\cdot\mathbf{a}^\prime_{j})$ with $\mathbf{a}^\prime_{1}=\mathbf{a}_{1}+\mathbf{a}_{2}$, and $\mathbf{a}^\prime_{2}=\mathbf{a}_{2}-\mathbf{a}_{1}$. The $+(-)$ sign refers to spin up (down) fermions. The eigenvalues of $H(\bk)=H_0(\bk)+H_{SO}(\bk)$ are $\epsilon_{\pm}(\bk)=\pm\sqrt{b_\bk+4\lambda_{SO}^2w^2_{\bk}}$ and $\epsilon_0(\bk)=0$ and their corresponding eigenstates can be written in the form,
\be
\left\vert u_{n\bk}\right\rangle =G_{n\bk}\left(
q_{A\bk},q_{B\bk},q_{C\bk}\right)^{\text{T}},
\label{eq:eigen}
\ee
where the expressions of the components $q_{l\bk}$ ($l=A,B,C$) and the normalization factor $G_{n\bk}$ for each band $n$($=0,\pm$) are given in Table~\ref{eigenstate_so}.

\begin{table}[th]
\caption{The expressions for the coefficients of $\left\vert u_{n\bk}\right\rangle$ with $x_{i}$=$\mathbf{k\mathtt{\cdot}a}_{i}$ ($i=1,2$). For $\bk=\mathbf{M}$, the listed expressions break down with ambiguity and hence we single them out, $|u_{+\mathbf{M}}\rangle=(0,1/\sqrt{2},i/\sqrt{2})^T$, $|u_{-\mathbf{M}}\rangle=(0,1/\sqrt{2},-i/\sqrt{2})^T$ and $|u_{0\mathbf{M}}\rangle=(1,0,0)^T$. Note that the upper (lower) sign in the square bracket refers to spin-up (down) fermions. $b_{\bk}=\sum_{i=1}^{2}[2t\cos(\bk\cdot \mathbf{a}_i)]^2$}
\label{eigenstate_so}
\begin{tabular}
[c]{cc}\hline\hline
$q_{A\bk}$ & $\ \ \ \ \ \ \ \ \ \ \ \ \ \ \ \ \ \ \epsilon_n^2(\bk)-4\lambda_{SO}^2w^2_{\bk}$\\
$q_{B\bk}$ & $\ \ \ \ \ \ \ \ \ \ \ \ \ \ \ \ \ \ 2t[\epsilon_n(\bk)\cos x_1
\pm 2i\lambda_{SO}w_{\bk}\cos x_2]$\\
$q_{C\bk}$ & $\ \ \ \ \ \ \ \ \ \ \ \ \ \ \ \ \ \ 2t[\epsilon_n(\bk)\cos x_2
\mp 2i\lambda_{SO}w_{\bk}\cos x_1]$\\
$G_{n\bk}^{-2}$ & $\ \ \ \ \ \ \ \ \ \ \ \ \ \ \ \ \ \epsilon_{n}^2(\bk)[\epsilon_n^2(\bk)-4\lambda_{SO}^2w_{\bk}^2
+b_{\bk}]$\\
 & \ \ \ \ \ \ \ \ \ \ \ \ \ \ \ \ \ $+4\lambda_{SO}^2w_{\bk}^2(b_{\bk}+4\lambda_{SO}^2w_{\bk}^2)$ \\\hline
\end{tabular}
\end{table}

At TRIM, we have eigenstates $|u_{-\mathbf{\Gamma}}\rangle=(\frac{1}{\sqrt{2}},\frac{-1}{2},\frac{-1}{2})^T$,
$|u_{-\mathbf{X}}\rangle=(\frac{1}{\sqrt{2}},0,\frac{1}{\sqrt{2}})^T$, $|u_{-\mathbf{Y}}\rangle=(\frac{1}{\sqrt{2}},\frac{-1}{\sqrt{2}},0)^T$, and $|u_{-\mathbf{M}}\rangle=(0,\frac{1}{\sqrt{2}},\frac{-i}{\sqrt{2}})^T$ according to Table~\ref{eigenstate_so}, respectively. Now, let us make a gauge transformation on eigenstates such that only one coordinate $\mathbf{R}$ would be assigned to each unit cell (instead of $\mathbf{R}+\mathbf{a}_i$ for each atom within the cell), and the Bloch Hamiltonian $H(\bk)$ now obeys $H(\bk+\mathbf{G})=H(\bk)$ and, with the inversion operator $\hat{P}$, $H(-\bk)=\hat{P}H(\bk)\hat{P}$, resulting in $[H(\bk=\Gamma_i),\hat{P}]=0$. Thus, the parity eigenvalue now becomes a good quantum number at TRIM. Under such transformation the eigenstates are rewritten as $|u_{-\mathbf{\Gamma}}\rangle=(\frac{1}{\sqrt{2}},\frac{-1}{2},
\frac{-1}{2})^T$ ,$|u_{-\mathbf{X}}\rangle=(\frac{1}{\sqrt{2}},0,\frac{1}{\sqrt{2}})^T$, $ |u_{-\mathbf{Y}}\rangle=(\frac{1}{\sqrt{2}},\frac{-1}{\sqrt{2}},0)^T$, and $|u_{-\mathbf{M}}\rangle=(0,\frac{-i}{\sqrt{2}},\frac{-1}{\sqrt{2}})^T$. Picking up any one of A-sites as our inversion center, we can determine the parity eigenvalue for each TRIM by examining the wave functions (eigenstates) in real space. Since moving a unit cell by $\mathbf{R}$ is simply multiplying our eigenstate at $\Gamma_i$ by a factor $e^{i\Gamma_i\cdot\bf{R}}$, we now see the parity eigenvalues for $\Gamma_i$=$(\mathbf{\Gamma},\mathbf{X},\mathbf{Y},\mathbf{M})$, are, respectively, $P$=$(+,+,+,-)$. The $Z_2$ index $\nu$ is then determined by
\be
(-1)^\nu=\prod_{i=1}^4\delta_i=-1,
\ee
where $\delta_i=\prod_{m=1}^{N}\xi_{2m}(\Gamma_i)$, and $N=1$ indicating either spin-up or spin-down band in our example here. Therefore, $\nu=1$ suggests that the ground state of our system at 1/3 filling has topological nature. We summarize the parity eigenvalue of each band in Table~\ref{parityev}. Finally, for $\varepsilon_A\neq 0$, one can follow similar procedure and the conclusion of $\nu=1$ is unchanged.

\begin{table}[th]
\caption{The parity eigenvalue for $n$-th band at TRIM $\Gamma_i$.}
\label{parityev}
\begin{tabular}
[c]{cccc}\hline\hline
$\Gamma_i$ & $n=+$ & $n=-$ & $n=0$ \\
$\mathbf{\Gamma}$ & + & + & +      \\
$\mathbf{X}$ & + & + & -           \\
$\mathbf{Y}$ & + & + & -           \\
$\mathbf{M}$ & - & - & +
\\\hline
\end{tabular}
\end{table}



\begin{thebibliography}{99}
\bibitem{klitzing86} See, for instance, K.~von Klitzing, Rev. Mod. Phys. {\bf 58}, 519 (1986); H.~L.~Stormer, D.~C.~Tsui, and A.~C.~Gossard, Rev. Mod. Phys. {\bf 71}, S298 (1999).
\bibitem{moore10} J.~E.~Moore, Nature {\bf 464}, 194 (2010).
\bibitem{rev} For recent reviews, see X.-L. Qi and S.-C. Zhang, arXiv:1008.2026; M.~Z.~Hasan and C.~L.~Kane, Rev. Mod. Phys. {\bf 82}, 3045 (2010).
\bibitem{kane05a} C.~L.~Kane and E.~J.~Mele, Phys. Rev. Lett. {\bf 95}, 146802 (2005).
\bibitem{kane05b} C.~L.~Kane and E.~J.~Mele, Phys. Rev. Lett. {\bf 95}, 226801 (2005).
\bibitem{bernevig06a} B.~A.~Bernevig and S.-C.~Zhang, Phys. Rev. Lett. {\bf 96}, 106802 (2006).
\bibitem{bernevig06b} B.~A.~Bernevig, T.~L.~Hughes, and S.-C.~Zhang, Science {\bf 314}, 1757 (2006).
\bibitem{fu07a} Liang Fu, C.~L.~Kane, and E.~J.~Mele, Phys. Rev. Lett. {\bf 98}, 106803 (2007).
\bibitem{fu07b} Liang Fu and C.~L.~Kane, Phys. Rev. B {\bf 76}, 045302 (2007).

\bibitem{wu04} C. Wu and S.-C. Zhang, Phys. Rev. Lett. {\bf 93}, 36403 (2004).

\bibitem{koenig07} M. Koenig, S. Wiedmann, C. Bruene, A. Roth, H. Buhmann, L. W. Molenkamp, X.-L. Qi, and S.-C. Zhang, Science {\bf 318}, 766 (2007).
\bibitem{hsieh08} D. Hsieh, D. Qian, L. Wray, Y. Xia, Y.~S.~Hor, R.~J.~Cava and M.~Z.~Hasan, Nature (London) {\bf 452}, 970 (2008).
\bibitem{hsieh09} D. Hsieh, Y. Xia, D. Qian, L. Wray, J. H. Dil, F. Meier, J. Osterwalder, L. Patthey, J. G. Checkelsky, N. P. Ong, A. V. Fedorov, H. Lin, A. Bansil, D. Grauer, Y. S. Hor, R.~J.~Cava, and M.~Z.~Hasan, Nature {\bf 460}, 1101 (2009).
\bibitem{xia09} Y. Xia, D. Qian, D. Hsieh, L. Wray, A. Pal, H. Lin, A. Bansil, D. Grauer, Y. S. Hor, R. J. Cava, and M.~Z.~Hasan, Nat. Phys. {\bf 5}, 398 (2009).
\bibitem{chen09} Y. L. Chen, J. G. Analytis, J.-H. Chu, Z. K. Liu, S.-K. Mo,
X.-L. Qi, H. J. Zhang, D. H. Lu, X. Dai, Z. Fang, S.-C. Zhang, I.~R.~Fisher, Z. Hussain, and Z.-X. Shen, Science {\bf 325}, 178 (2009).

\bibitem{scz08} S. Raghu, X.-L. Qi, C. Honerkamp, and S.-C.~Zhang, Phys. Rev. Lett. {\bf 100}, 156401 (2008).
\bibitem{weeks10a} C.~Weeks and M.~Franz, Phys. Rev. B {\bf 81}, 085105 (2010).
\bibitem{xu10} C. Xu, Phys. Rev. B {\bf 83}, 024408 (2011).

\bibitem{yingran} K.-Y. Yang, W. Zhu, D. Xiao, S. Okamoto, Z. Wang, and Y. Ran, Phys. Rev. B {\bf 84}, 201104(R) (2011).
\bibitem{fiete2} A. Ruegg and G. A. Fiete, Phys. Rev. B~ {\bf 84}, 201103(R) (2011).
\bibitem{fiete3} A. R¨¹egg, C. Mitra, A. A. Demkov, and G. A. Fiete, Phys. Rev. B {\bf 85}, 245131 (2012). 
\bibitem{fiete4} A. R¨¹egg, C. Mitra, A. A. Demkov, and G. A. Fiete, Phys. Rev. B {\bf 88}, 115146 (2013).

\bibitem{sun09} K.~Sun, H. Yao, E.~Fradkin and S.~A.~Kivelson, Phys. Rev. Lett. {\bf 103}, 046811 (2009).
\bibitem{liu10} Q.~Liu, H. Yao, and Tianxing~Ma, Phys. Rev. B {\bf 82}, 045102 (2010).
\bibitem{fiete1} J. Wen, A. R¨¹egg, C.-C. J. Wang, and G. A. Fiete, Phys. Rev. B {\bf 82}, 075125 (2010). 

\bibitem{oskar14} J. M. Murray and O. Vafek, Phys. Rev. B 89, 201110(R).
\bibitem{zhang09} Y. Zhang, Y. Ran, and Ashvin Vishwanath, Phys. Rev. B {\bf 79}, 245331 (2009).
\bibitem{vafek10} O.~Vafek and Kun Yang, Phys. Rev. B {\bf 81}, 041401(R) (2010).
\bibitem{weeks10b} C.~Weeks and M.~Franz, Phys. Rev. B {\bf 82}, 085310 (2010).
\bibitem{green10} D.~Green, L.~Santos, and C.~Chamon, Phys. Rev. B {\bf 82}, 075104 (2010).
\bibitem{oganesyan01} V.~Oganesyan, S.~A.~Kivelson, and E.~Fradkin, Phys. Rev. B {\bf 64}, 195109 (2001).
\bibitem{kivelson03} S.~A.~Kivelson, I.~P.~Bindloss, E.~Fradkin, V.~Oganesyan, J.~M. Tranquada, A.~Kapitulnik, and C.~Howald, Rev. Mod. Phys. {\bf 75}, 1201 (2003).
\bibitem{wu07} C.~Wu, K.~Sun, E.~Fradkin, and S.-C.~Zhang, Phys. Rev. B {\bf 75}, 115103 (2007).
\bibitem{chong08} Y.~D.~Chong, X.-G.~Wen, and M.~Soljacic, Phys. Rev. B {\bf 77}, 235125 (2008).
\bibitem{sun08} Kai~Sun and E.~Fradkin, Phys. Rev. B {\bf 78}, 245122 (2008).
\bibitem{uebelacker11} S.~Uebelacker and C.~Honerkamp, Phys. Rev. B {\bf 84}, 205122 (2011).
\bibitem{haldane04} F.~D.~M. Haldane, Phys. Rev. Lett. {\bf 93}, 206602 (2004).
\bibitem{bergman08} D.~L.~Bergman, Congjun Wu, and L.~Balents, Phys. Rev. B {\bf 78}, 125104 (2008).
\bibitem{goldman11} N.~Goldman, D.~F.~Urban, and D.~Bercioux, 	Phys. Rev. A {\bf 83}, 063601 (2011).
\bibitem{shen11} R.~Shen, L.~B.~Shao, Baigeng Wang, and D.~Y.~Xing, Phys. Rev. B {\bf 81}, 041410(R) (2010).
\bibitem{nielssen83} H. Nielssen and N. Ninomiya, Phys. Lett. {\bf 130B}, 389
(1983).
\bibitem{haldane88} F.~D.~M. Haldane, Phys. Rev. Lett. {\bf 61}, 2015 (1988).
\bibitem{varma06} C.~M.~Varma, Phys. Rev. B {\bf 73}, 155113 (2006).
\bibitem{particlehole} If one replaces $n_i$ by $n_i-\rho_0$ with $\rho_0$ denoting the average charge density and rewrites the density-density repulsions in terms of it in Eq.~(\ref{eq:int}), the mapping between 1/3 and 2/3 fillings by particle-hole language translation will become exact.
\bibitem{FM} A ferromagnetic (FM) state can be proved to be the exact ground state in systems with half-filled flat bands and short-range repulsions. However, in our system the flat band (which in fact acquires a small band width if a small $t^{\prime\prime}$ exists) is completely filled (or empty), and thus it is the dynamics around the BCP, rather than the filled band, that determines the phase. In addition, we notice that the FM state cannot open a full gap on the Fermi surface. Instead, it results in a (circular) line node with finite DOS at Fermi level, in sharp contrast to a typical nematic state, where there are only discrete Dirac nodes. Therefore, it is never favored energetically.
\bibitem{lieb89} E.~H.~Lieb, Phys. Rev. Lett. {\bf 62}, 1201 (1989).
\bibitem{emery87} V.~J.~Emery, Phys. Rev. Lett. {\bf 58}, 2794 (1987).
\bibitem{dagotto94} E.~Dagotto, Rev. Mod. Phys. {\bf 66}, 763 (1994) and references therein.
\bibitem{thomale08} R.~Thomale and M.~Greiter, Phys. Rev. B {\bf 77}, 094511 (2008).
\bibitem{hybertsen89} M.~S.~Hybertsen, M.~Schluter, and N.~E.~Christensen,
Phys. Rev. B {\bf 39}, 9028 (1989).
\bibitem{sun10} Kai Sun, W. Vincent Liu, and S.~Das Sarma, Nature Physics 8, 67 (2012).
\bibitem{tang11} E. Tang, J.-W. Mei, and X.-G. Wen, Phys. Rev. Lett. {\bf 106}, 236802 (2011).
\bibitem{neupert11} T. Neupert, L. Santos, C. Chamon, and C. Mudry
Phys. Rev. Lett. {\bf 106}, 236804 (2011).
\bibitem{sun11} K. Sun, Z.-C.~Gu, H.~Katsura, and S.~Das Sarma, Phys. Rev. Lett. {\bf 106}, 236803 (2011).
\bibitem{sheng11} D. N. Sheng, Z.-C.~Gu, Kai Sun, and L. Sheng, Nature Communications 2, 389 (2011).
\bibitem{bernevig11} N. Regnault and B. Andrei Bernevig, Phys. Rev. X {\bf 1}, 021014 (2011); Yang-Le Wu, B. Andrei Bernevig, and N. Regnault, Phys. Rev. B {\bf 85}, 075116 (2012).
\bibitem{lu11} Y.-M. Lu and Y. Ran, Phys. Rev. B {\bf 85}, 165134 (2012).
\bibitem{shankar94} R.~Shankar, Rev. Mod. Phys. {\bf 66}, 129 (1994).

\end{thebibliography}
\end{document}